\documentclass[aps,prb,reprint,superscriptaddress]{revtex4-1}

\usepackage{chemformula}
\usepackage{graphicx}
\usepackage{xspace}
\usepackage{bbm}
\usepackage{multirow}
\usepackage{tabularx}
\usepackage{adjustbox}
\usepackage{placeins}
\usepackage{url}
\usepackage{mathtools}
\usepackage{bm}
\usepackage{multirow}
\usepackage{algorithm}
\usepackage{algpseudocode}
\newcommand*{\Continue}{\textbf{continue}}

\usepackage{amssymb,amsmath}

\newcommand{\set}[1]{\ensuremath{\mathbf{#1}}}

\DeclareMathOperator{\rank}{rank}
\newcommand{\calcrank}[1]{\rank\left( #1 \right)}
\DeclareMathOperator\lcm{lcm}

\usepackage{array}
\newcolumntype{L}[1]{>{\raggedright\let\newline\\\arraybackslash\hspace{0pt}}m{#1}}
\newcolumntype{C}[1]{>{\centering\let\newline\\\arraybackslash\hspace{0pt}}m{#1}}
\newcolumntype{R}[1]{>{\raggedleft\let\newline\\\arraybackslash\hspace{0pt}}m{#1}}

\newcommand{\Bstrut}[1]{\rule[- #1 ex]{0pt}{0pt}}   

\usepackage{fancyhdr}
\fancypagestyle{arxivfooter}{%
  \fancyhf{}
  
  \fancyfoot[L]{Published in Phys. Rev. Materials \textbf{3}, 034003 (2019) \hfill \textcopyright 2019 American Physical Society\\ \url{http://doi.org/10.1103/PhysRevMaterials.3.034003}}
}

\begin{document}


\title{Definition of a scoring parameter to identify low-dimensional materials components.}



\author{Peter Mahler Larsen}
\email{pmla@mit.edu}
\affiliation {Center for Atomic-scale Materials Design (CAMD), Department of Physics, Technical University of Denmark, 2800 Kongens Lyngby, Denmark}
\affiliation {Department of Materials Science and Engineering, MIT, Cambridge, MA 02139, USA}
\author{Mohnish Pandey}
\author{Mikkel Strange}
\author{Karsten Wedel Jacobsen}
\affiliation {Center for Atomic-scale Materials Design (CAMD), Department of Physics, Technical University of Denmark, 2800 Kongens Lyngby, Denmark}

\date{\today}

\begin{abstract}
The last decade has seen intense research in materials with reduced
dimensionality. The low dimensionality leads to interesting electronic
behavior due to electronic confinement and reduced screening. The
investigations have to a large extent focused on 2D materials both in
their bulk form, as individual layers a few atoms thick, and through
stacking of 2D layers into heterostructures. The identification of
low-dimensional compounds is therefore of key interest.
Here, we perform a geometric analysis of material structures, demonstrating
a strong clustering of materials depending on their dimensionalities.
Based on the geometric analysis, we propose a simple scoring parameter to identify
materials of a particular dimension or of mixed dimensionality. The
method identifies spatially connected components of the materials and
gives a measure of the degree of ``1D-ness,'' ``2D-ness,'' etc., for
each component. The scoring parameter is applied to the
Inorganic Crystal Structure Database and the Crystallography Open Database
ranking the materials according to their degree of dimensionality. In
the case of 2D materials the scoring parameter is seen to clearly
separate 2D from non-2D materials and the parameter correlates well
with the bonding strength in the layered materials. About 3000
materials are identified as one-dimensional, while more than 9000
are mixed-dimensionality materials containing a molecular (0D) component.
The charge states of the components in selected highly ranked materials are investigated using
density functional theory and Bader analysis showing that the
spatially separated components have either zero charge, corresponding
to weak interactions, or integer charge, indicating ionic bonding.
\end{abstract}


\maketitle

\thispagestyle{arxivfooter}

\section{Introduction}
Low-dimensional materials with one or more characteristic lengths of
the materials limited to the atomic scale have received significant
attention recently. Since the discovery of graphene the world has seen
intense research in 2D materials involving synthesis and investigation
of mechanical, electronic, magnetic, and catalytic properties of new
materials \cite{Novoselov:2004it, Bhimanapati:2015bo, Ferrari:2015co,
  Zeng:2018bz}. Also a number of computational efforts have been
dedicated to the identification of new 2D materials and to the
construction of computational databases with information about their
stability and (photo-) electronic properties \cite{Mounet:2018ks,
  ashton2017topology, Haastrup:2018vr}. One of the driving forces
behind this research has been an interest in ultra-small electronic
components and this has also led to studies of 1D or quasi-1D
materials as possible interconnects \cite{Stolyarov:2016jy,
  Geremew:2018fv}. Furthermore, the possibility of combining materials
of different dimensionality into new van der Waals bonded
mixed-dimensional heterostructures has recently been discussed
\cite{Jariwala:2017el}. The realization of such structures relies on the
identification of appropriate weakly interacting material components
of different dimensionalities.

In the following we shall define a simple geometrical scoring
parameter to identify low-dimensional components in existing
materials. The scoring parameter is easy to compute and can be applied
to large materials databases. We illustrate this by mining the
Inorganic Crystal Structure Database~\cite{bergerhoff1983icsd} (ICSD)
and the Crystallography Open Database~\cite{grazulis2012cod} (COD)
to find materials with clearly identifiable low-dimensional atomic
structures. The identified materials consist of weakly interacting
components as we demonstrate for 2D materials by comparison with
previously calculated exfoliation energies. Apart from being
interesting in their own right, the materials components may also form
templates for substitution of similar chemical elements to form new
materials of different dimensions \cite{Mounet:2018ks,
  Haastrup:2018vr}.

\section{Results and discussion}

\subsection{Bond-length interval analysis}

The definition of the scoring parameter requires first, that we can
identify the dimension(s) of a periodic solid.  Given an atom in a
bonded cluster, the cluster dimension is given by the rank of the
subspace spanned by the atom and its periodically connected
neighbors.  We refer to this method as the rank determination
algorithm (RDA), which is described in detail in the Methods section.

\def\arrowtextA{\leaders\vrule height2.7pt depth-2.3pt\hfil}
\def\arrowtext#1#2{\hbox to#1{\arrowtextA\ #2 \arrowtextA\kern2pt\llap{$\succ$}}}

\begin{figure*}
\begin{minipage}{0.93\columnwidth}
  \centering
  \textbf{(a)}\vspace{2mm}\\
  \includegraphics[width=0.8\columnwidth]{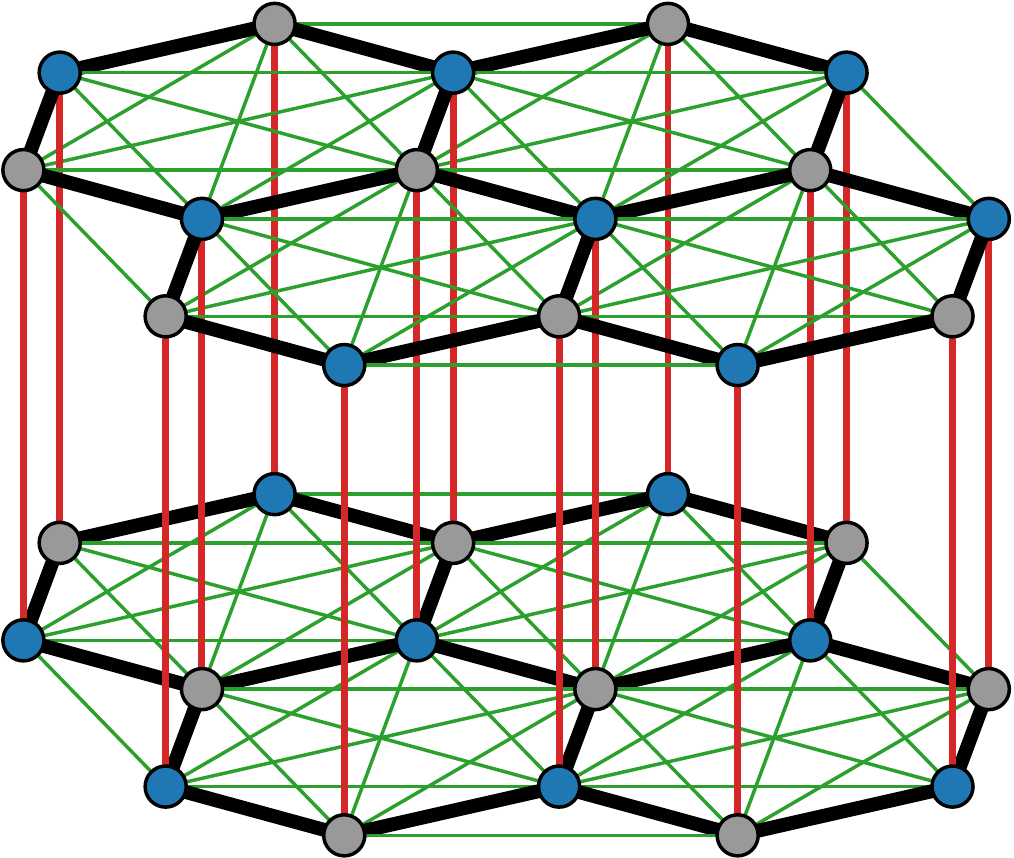}
\end{minipage}
\hspace{3mm}
\begin{minipage}{0.93\columnwidth}
  \centering
  \textbf{(b)}\vspace{6.5mm}\\
  Boron Nitride (BN)\vspace{1mm}\\
  \includegraphics[width=0.9\columnwidth]{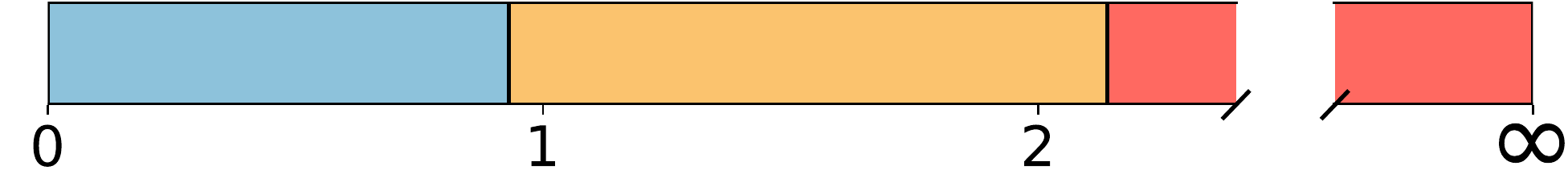}\\
  $\arrowtext{3.5cm}{\textbf{\emph{k}}}$\vspace{4mm}\\
  \ch{((CH3)2 NH2)2 (Al2H(PO4)3)}\vspace{1mm}\\
  \includegraphics[width=0.9\columnwidth]{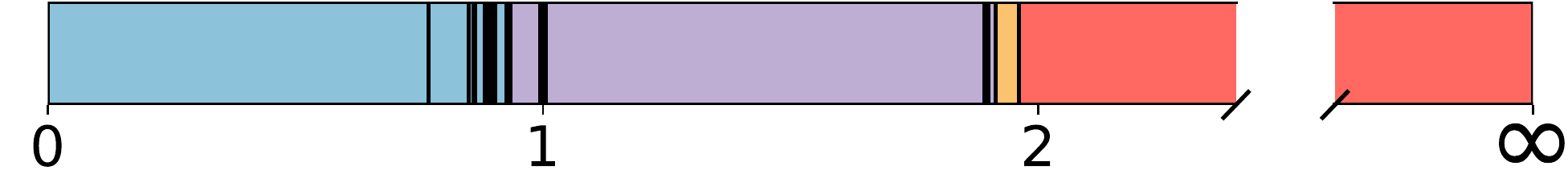}\\
  $\arrowtext{3.5cm}{\textbf{\emph{k}}}$\vspace{5mm}\\
  \hspace{3mm}\textbf{Legend:}\hspace{2mm}
\includegraphics[width=0.06\columnwidth]{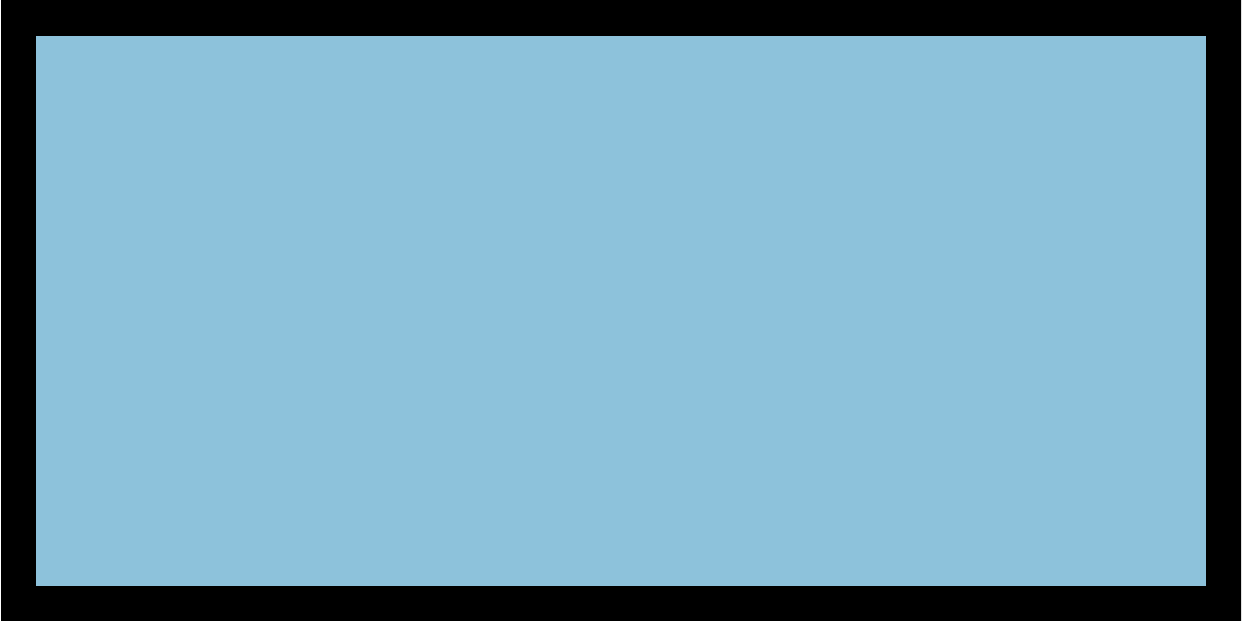} \textbf{0D}\hspace{2mm}
\includegraphics[width=0.06\columnwidth]{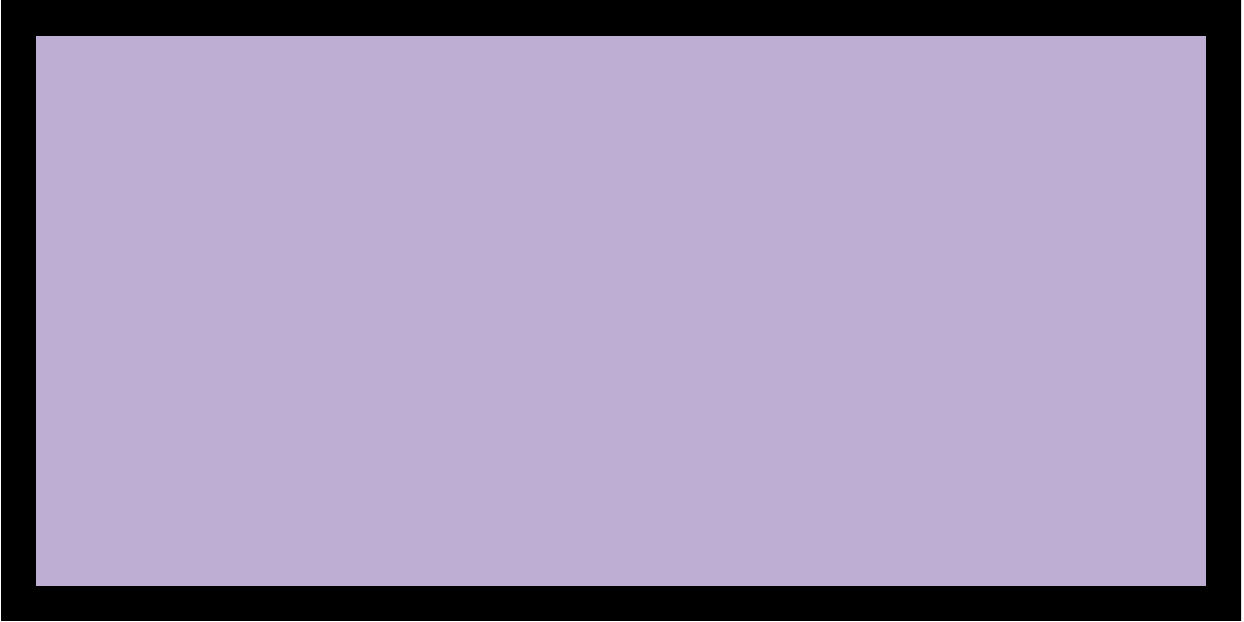} \textbf{02D}\hspace{2mm}
\includegraphics[width=0.06\columnwidth]{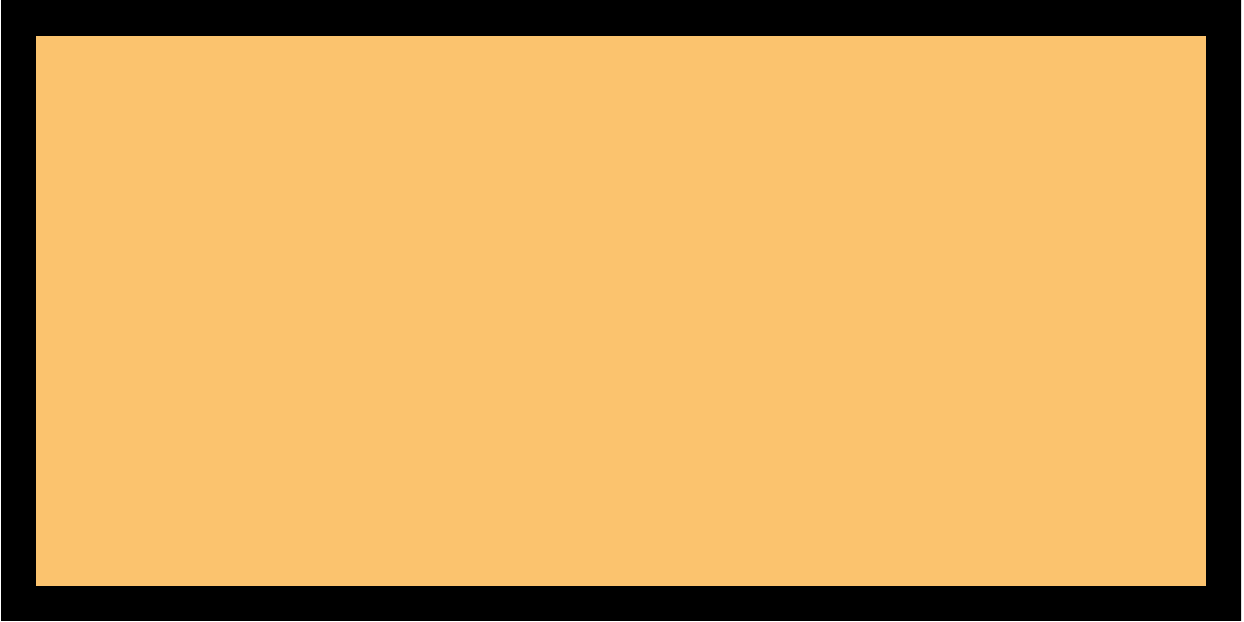} \textbf{2D}\hspace{2mm}
\includegraphics[width=0.06\columnwidth]{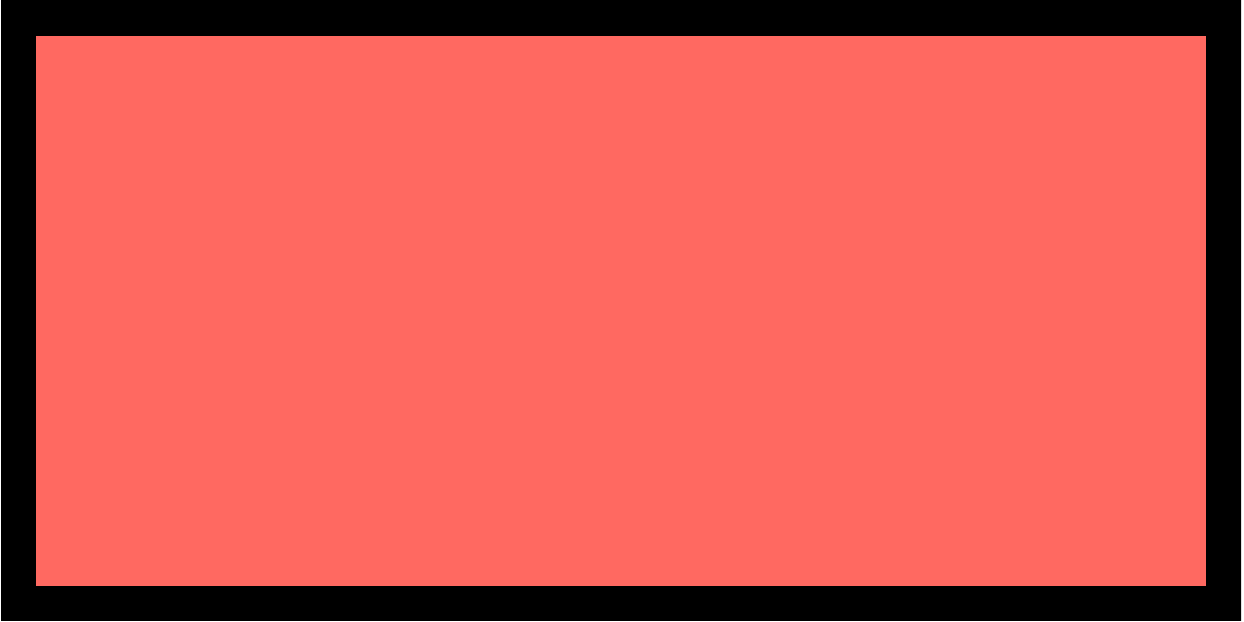} \textbf{3D}
  \vspace{2mm}\\
\end{minipage}
\caption{
\textbf{(a)}
Boron nitride (BN) in a layered structure. Edges are
colored according to their effect on the dimensionality
classification. Black edges are the strong covalent bonds, which
result in a 2D classification. Green edges are longer
bonds, which do not change the classification from 2D. Red edges are
weak bonds which result in a 3D classification.
\textbf{(b)}
Illustration of the change in dimensionality classification
with increasing $k$, for boron nitride and a layered
aluminophosphate structure with intercalated organic molecules.
Larger values of $k$ increase the dimensionality. Multiple intervals
with the same dimensionality can exist, though these have different
numbers of components. The best classification corresponds to a wide
interval starting at $k \approx 1$.
\label{fig:kinterval}
}
\end{figure*}

An accurate identification of bonded clusters requires a full
electronic structure calculation, where
the bond strength and character can be addressed. However, for purposes
of screening large materials databases this approach is computationally
infeasible. Instead, we use a simple geometric criterion for bonding.
We describe two atoms, $i$ and $j$, as bonded if the distance between
them is less than a specified multiple of their covalent radius sum:
\begin{equation}
  \label{eq:bond}
d_{ij} < k \left(r^{\text{cov}}_i + r^{\text{cov}}_j\right),  
\end{equation}
Here, $d_{ij}$ is the distance between atoms $i$ and $j$,
$r^{\text{cov}}_i$ and $r^{\text{cov}}_j$ are the corresponding
covalent radii~\cite{cordero2008covalent}, and $k$ is a variable to be investigated.
The latter choice is motivated by the strong dependence of the
classification of the dimensionality of a material upon the $k$ value;
as illustrated for the boron-nitride structure in
Fig.~\ref{fig:kinterval}(a), too small a $k$ value will
underestimate the dimensionality, whereas too large a $k$ value will
overestimate it. Rather than attempt to identify a good value of $k$,
we observe that, for any given structure, there exists a finite number
of relevant $k$ intervals to investigate.

We start by considering the set of interatomic distances in a
material, sorted by increasing $k$ value [where
$k = (r^{\text{cov}}_i + r^{\text{cov}}_j) / d_{ij}$]. Each interatomic
distance corresponds to a possible bond; as shown in
Fig.~\ref{fig:kinterval}(a), bonds can be physical or not. Bonds are
inserted one at a time, and at each step the RDA is used to
determine the number of components and their
dimensionality. Initially, every atom is a separate 0D component; as
more bonds are inserted, the number of components decreases and the
component dimensionalities increase. The process terminates when a
single 3D component is left; \textit{i.e.}, all atoms are contained in the same
bonded cluster. This process finds all $k$ value
intervals in which the dimensionality classification is constant, of which there are a finite number.
The interval identification method is described in more detail in the
Methods section.

Fig.~\ref{fig:kinterval}(b) shows the application of this method to two
different layered structures. It can be seen that different
dimensionality classifications exist at different $k$ values.
Furthermore, the intervals have very different widths;
the first interval
is of the form
$[0, k)$, whereas the last interval
is of the form $[k, \infty)$.

\begin{figure*}
\begin{minipage}{\columnwidth}
  \centering
  \textbf{(a)}\vspace{2mm}\\
  \includegraphics[width=\columnwidth]{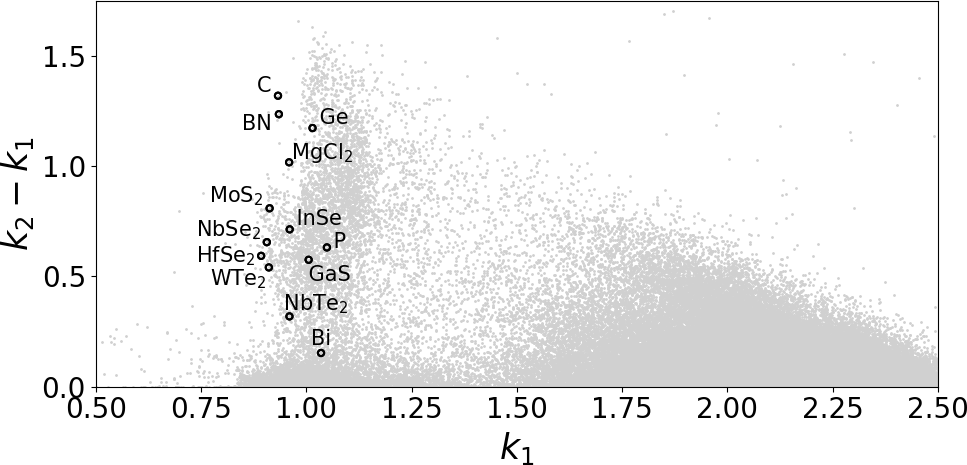}
\end{minipage}
\hspace{3mm}
\begin{minipage}{\columnwidth}
  \centering
  \textbf{(b)}\vspace{2mm}\\
  \includegraphics[width=\columnwidth]{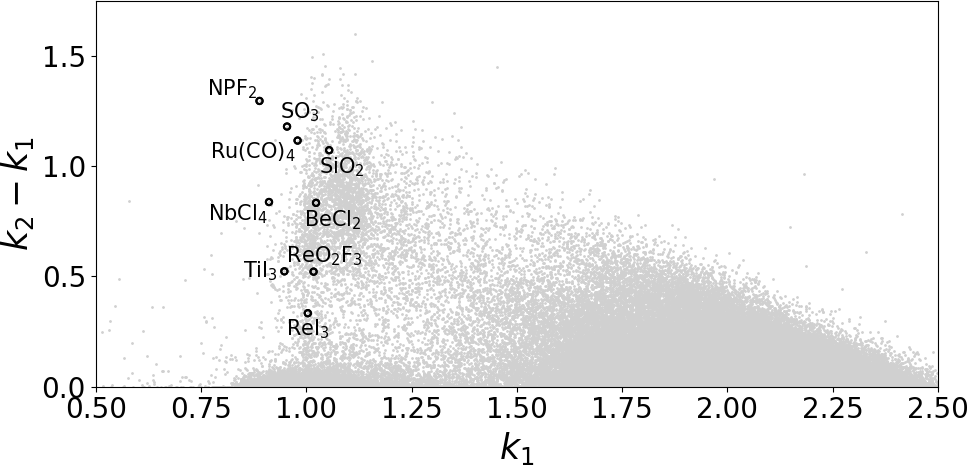}
\end{minipage}
\\
\begin{minipage}{\columnwidth}
  \centering
  \textbf{(c)}\vspace{2mm}\\
  \includegraphics[width=\columnwidth]{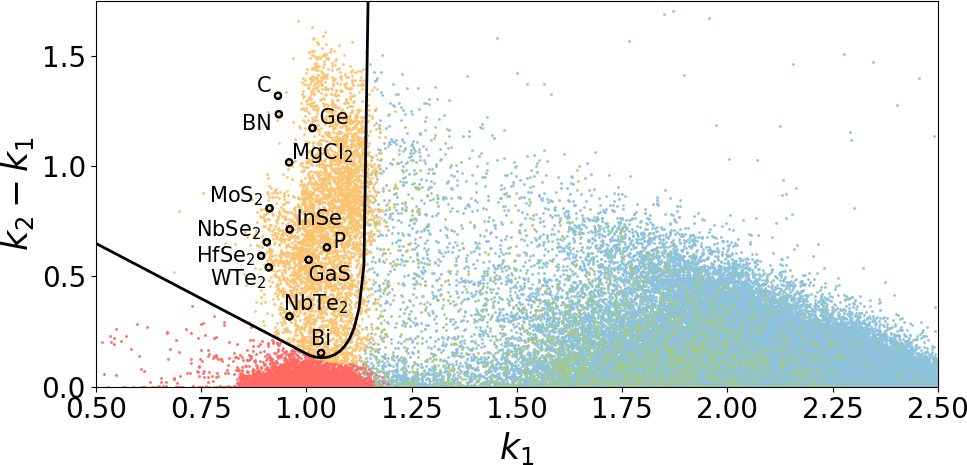}
\end{minipage}
\hspace{3mm}
\begin{minipage}{\columnwidth}
  \centering
  \textbf{(d)}\vspace{2mm}\\
  \includegraphics[width=\columnwidth]{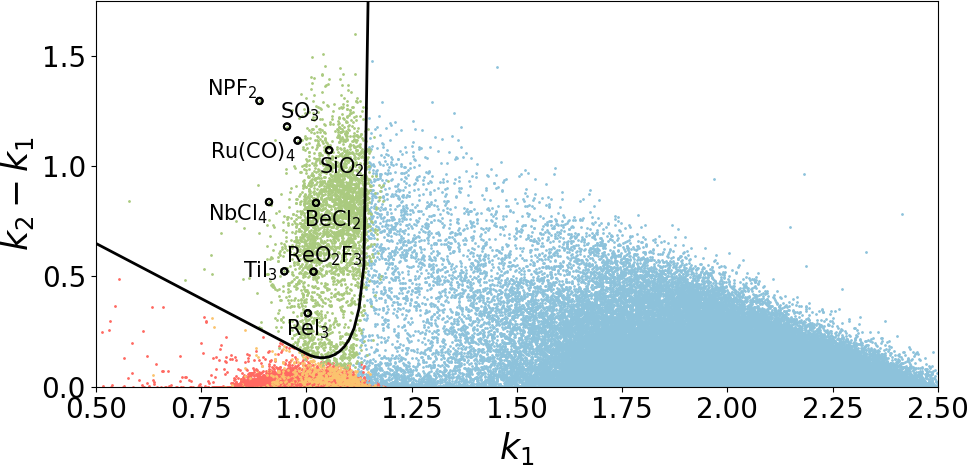}
\end{minipage}
\\
\includegraphics[width=0.06\columnwidth]{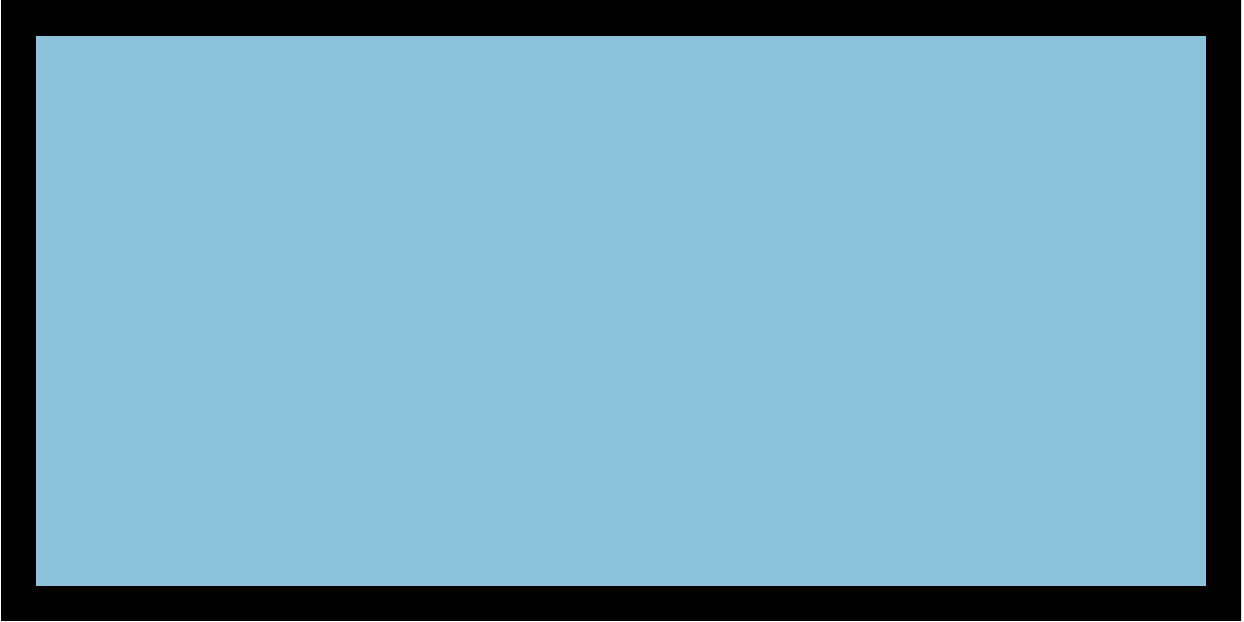} \textbf{0D}\hspace{3mm}
\includegraphics[width=0.06\columnwidth]{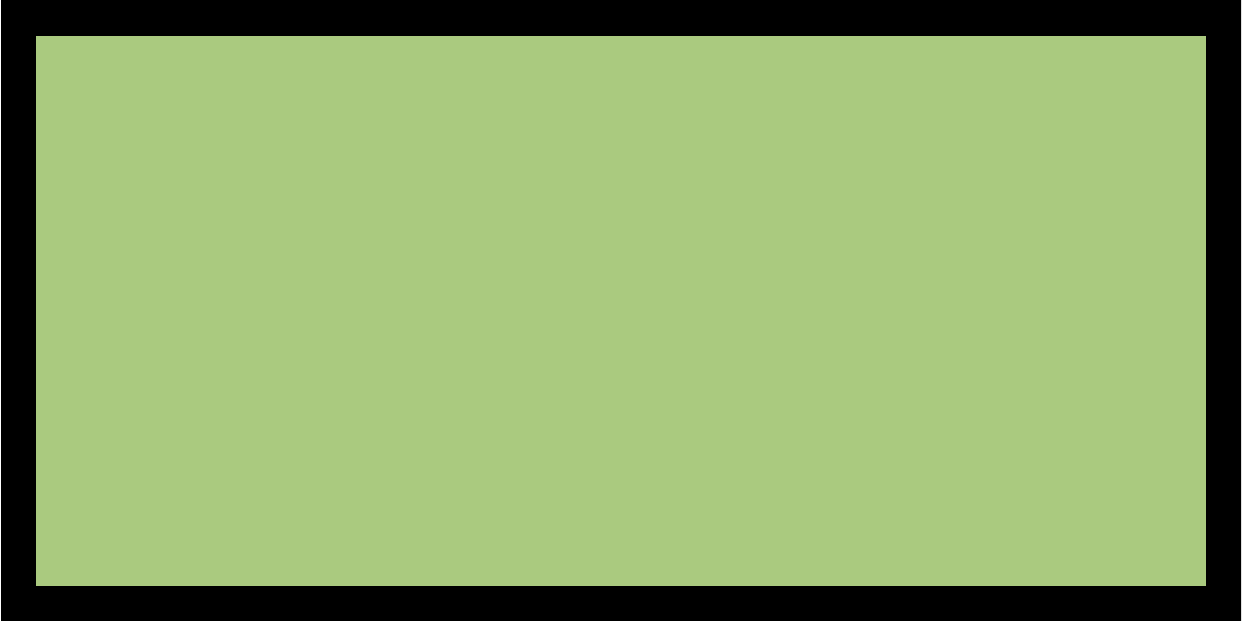} \textbf{1D}\hspace{3mm}
\includegraphics[width=0.06\columnwidth]{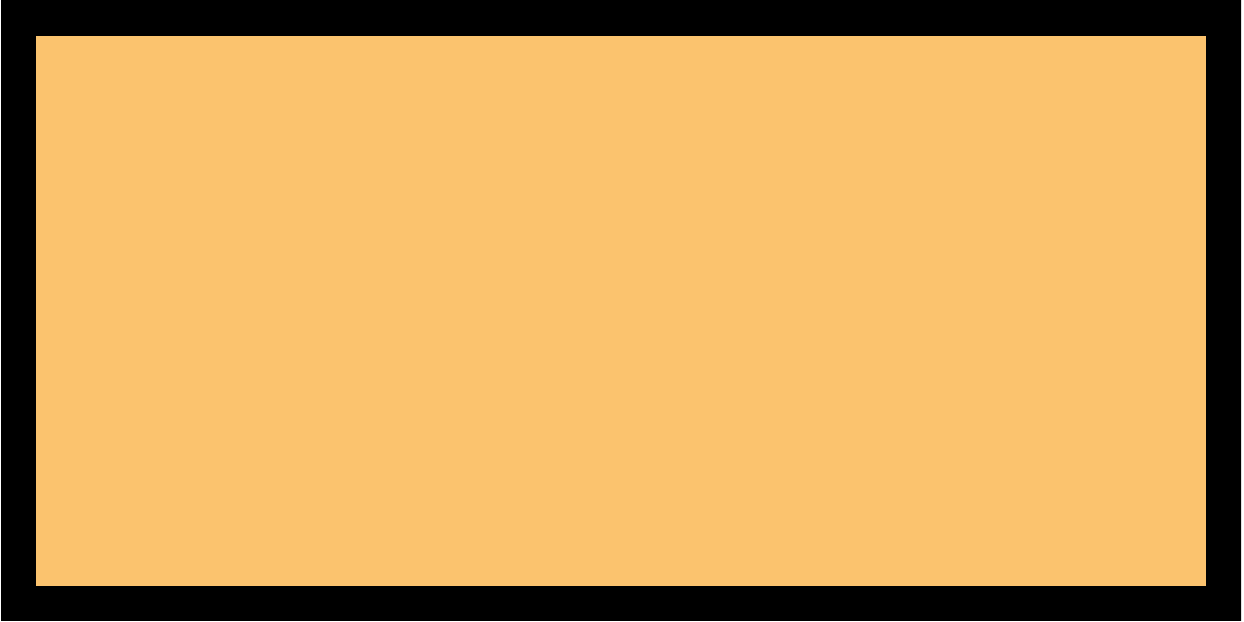} \textbf{2D}\hspace{3mm}
\includegraphics[width=0.06\columnwidth]{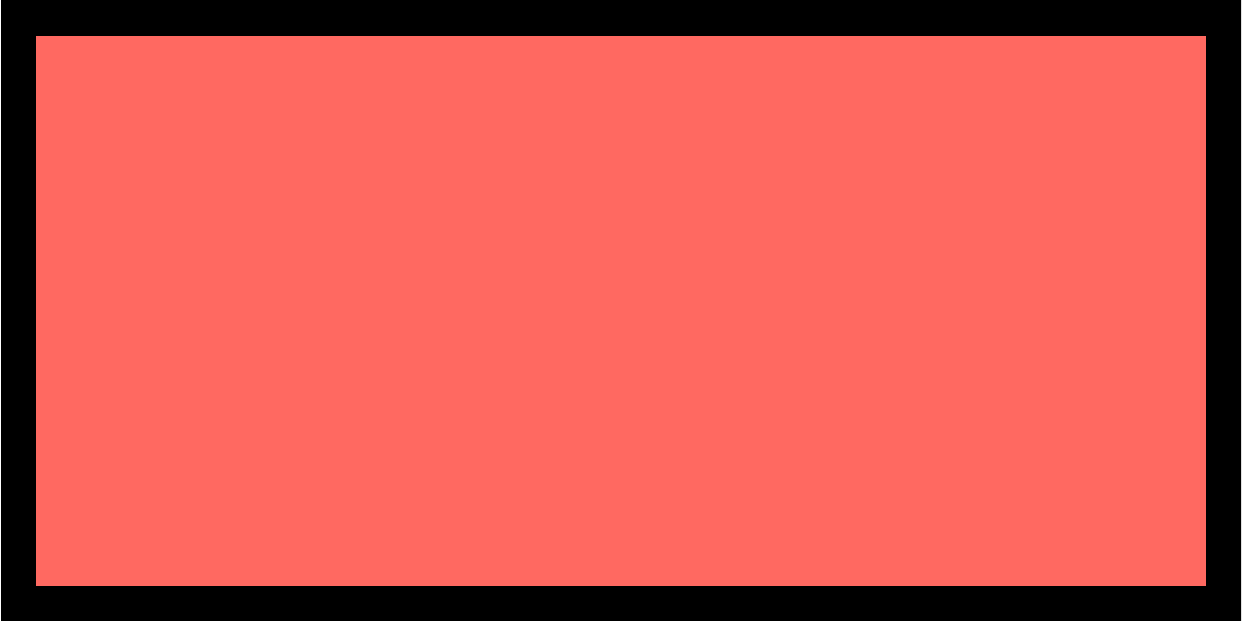} \textbf{3D}
\caption{
\textbf{(a)}
Interval plot of all structures in the ICSD and COD with a 2D interval, with some well-known
structures marked.
\textbf{(b)}
Same as (a) but for 1D intervals.
\textbf{(c)}, \textbf{(d)}
Same as (a), (b) but with intervals colored according to dimensionality using the scoring parameter.  For clarity, mixed-dimensionality structures are not shown.  The line shows the contour $s \left( k_1, k_2 \right) = 0.5$.
\label{fig:clustering}
}
\end{figure*}

\subsection{Defining the scoring parameter}

Figs.~\ref{fig:clustering}(a) and \ref{fig:clustering}(b) show the $k$ intervals for all structures in the ICSD and COD with, respectively, a 2D interval and a 1D interval.
In both cases there is a visible cluster of structures in the approximate region $k_1 \approx 1$ and $0.1 \leq k_2 - k_1 \leq 1.5$.  
The position of the cluster is intuitive from a bonding perspective.  First, 
if the bonding model and covalent radii exactly described the actual bond lengths, the cluster would lie on the line $k_1=1$; the variability in the interval starting points results from the simplicity of the ball-and-stick bonding model.  Second, since low-dimensional components are geometrically separated, we should expect a correspondingly wide $k$ interval; it can be seen that easily exfoliable structures such as graphite, boron nitride, and molybdenum disulfide have wide $k$ intervals.

\begin{figure}
  \centering
  \includegraphics[width=8cm]{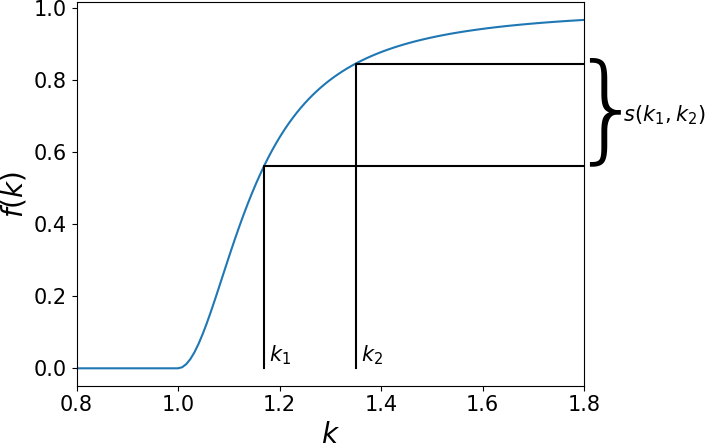}
  \caption{Variation of $f\left( k \right)$ versus $k$ and the dependence of the scoring parameter on the $k$ interval.}
  \label{fig:dimscore}
\end{figure}

We propose a scoring parameter which distills the above observations of the $k$ interval plots into a single number:
\begin{equation}
s\left(k_1, k_2 \right) = f\left(k_2 \right) - f\left(k_1 \right)
\label{eq:scoring_scheme_s}
\end{equation}
where
\begin{equation}
f\left(x\right) = \frac{ c \times \max(0, x-1)^2 }{1 + c \times \max(0, x-1)^2}
\label{eq:scoring_scheme_f}
\end{equation}
Here, $c$ is a constant which determines the scale at which a bond is
broken.  We use $c = 1 / 0.15^2$, which is chosen so that
$s \left( 1, 1.15 \right) = s \left( 1.15, \infty \right) = 0.5$;
 slightly different values of the parameter will give similar results.
Fig.~\ref{fig:dimscore} illustrates how a $k$ interval is
transformed into a score.

The interval width increases the score, but with
diminishing returns as $k_1$ increases above 1.  This avoids the
$\left[k, \infty \right)$ interval dominating unless $k_1$ is close to
1, in which case the structure is indeed 3D.  Furthermore, $k$ values
below 1 are effectively set to 1; this avoids erroneous
low-dimensional classifications when
$\left[k_1, k_2\right] \approx [0, 1]$.
In structures with low-dimensional components, the scoring parameter rewards large intercomponent distances.
A further convenient property of the scoring scheme is that the interval scores sum to 1.  We have found that the best results are achieved by merging $k$ intervals with the same types of dimensionalities [e.g.~intervals of the same color in Fig.~\ref{fig:kinterval}(b)].

The principal motivation of the scoring scheme is to identify the intuitively correct dimensionality classification, by determining whether a $k$ interval lies within a cluster of the type shown in
Figs.~\ref{fig:clustering}(a) and \ref{fig:clustering}(b).  Using the scoring parameter, the structures are colored in Figs.~\ref{fig:clustering}(c) and \ref{fig:clustering}(d) according to their dimensionality classification.

\begin{figure*}
{
\setlength{\fboxsep}{0pt}%
\centering
\begin{tabular}{ccc}
\footnotesize{COD 2300448}
&\footnotesize{COD 9011286}
&\footnotesize{COD 411179}\\
\fbox{\includegraphics[trim={0 2cm 0 1.7cm},clip, width=0.3\textwidth]{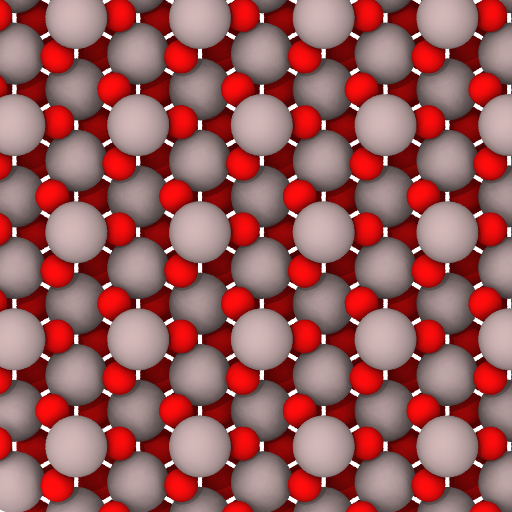}}
& \fbox{\includegraphics[trim={0 2.2cm 0 1.5cm},clip, width=0.3\textwidth]{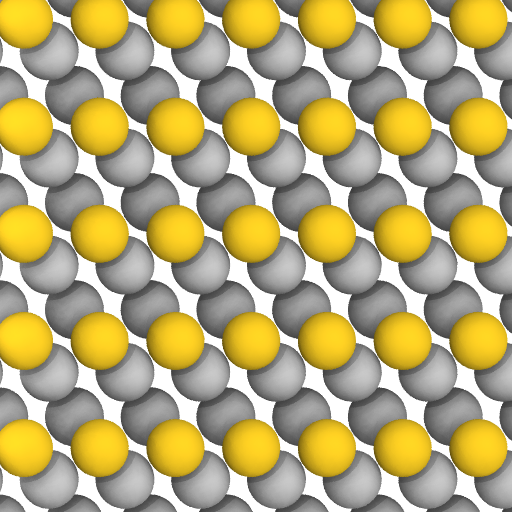}}
& \fbox{\includegraphics[trim={0 2cm 0 1.7cm},clip, width=0.3\textwidth]{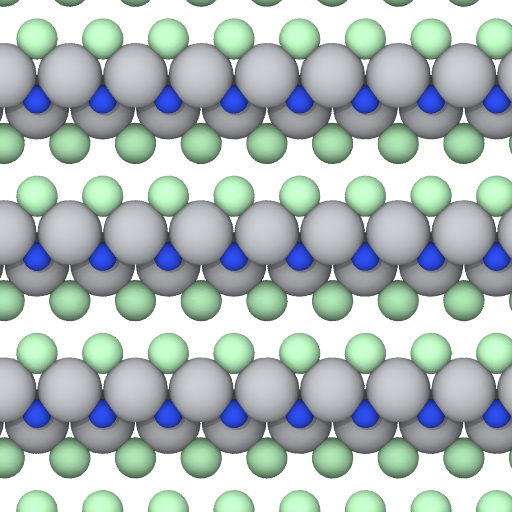}}\\
\ch{Al2O3}
& \ch{AuTe2}
& \ch{Ti2Cl2N2}
\\
  $s_{2\text{D}} =$ \makebox[0pt][l]{0.12}\phantom{99999} $k = [$ \makebox[0pt][l]{$0.99,$}\phantom{9999} \makebox[0pt][l]{$1.05$}\phantom{9999} $]$
& $s_{2\text{D}} =$ \makebox[0pt][l]{0.47}\phantom{99999} $k = [$ \makebox[0pt][l]{$1.02,$}\phantom{9999} \makebox[0pt][l]{$1.14$}\phantom{9999} $]$
& $s_{2\text{D}} =$ \makebox[0pt][l]{0.97}\phantom{99999} $k = [$ \makebox[0pt][l]{$0.93,$}\phantom{9999} \makebox[0pt][l]{$1.81$}\phantom{9999} $]$
\\
  $s_{3\text{D}} =$ \makebox[0pt][l]{0.88}\phantom{99999} $k = [$ \makebox[0pt][l]{$1.05,$}\phantom{9999} \makebox[0pt][l]{$\;\infty$}\phantom{9999} $]$
& $s_{3\text{D}} =$ \makebox[0pt][l]{0.52}\phantom{99999} $k = [$ \makebox[0pt][l]{$1.14,$}\phantom{9999} \makebox[0pt][l]{$\;\infty$}\phantom{9999} $]$
& $s_{3\text{D}} =$ \makebox[0pt][l]{0.03}\phantom{99999} $k = [$ \makebox[0pt][l]{$1.81,$}\phantom{9999} \makebox[0pt][l]{$\;\infty$}\phantom{9999} $]$
\\
\end{tabular}
\caption{Structures with successively larger interlayer spacings. All
  three structures contain intervals for both 2D and 3D
  classifications.  The scoring scheme suggests the most likely dimensionality classification.
  In cases where multiple reasonable	classifications are possible, as in the \ch{AuTe2} structure,
  the ambiguity is reflected in the scoring scheme.}
\label{fig:successive}
}
\end{figure*}

The scoring scheme is demonstrated for three structures in
Fig.~\ref{fig:successive}.  The first material, \ch{Al2O3}, is
clearly a bulk crystalline structure.  If a single $k$ value threshold
at $k \approx 1$ were used, however, it would result in a
misclassification as a layered structure.  Similarly, the scoring
scheme also ensures that the \ch{Ti2Cl2N2} structure is correctly
identified as a layered structure.  The \ch{AuTe2} has an ambiguous classification,
lying close to the contour $s_2 \left( k_1, k_2 \right) = 0.5$.  In this case the
dimensionality classification is sensitive to small changes in the 
functional form or the parameters of the scoring function.  Then, the useful information
contained in the scores is not in their exact values, but rather that $s_2$ and $s_3$ are approximately
equal in value; this can be interpreted as a layered structure with a very small interlayer spacing.

It should be emphasized that the scoring is exclusively based
on interatomic distances and atomic sizes, and that it simply assumes
that longer bonds tend to be weaker than shorter ones. The physical
characters of the bonds, \textit{i.e.}\ whether they can be considered covalent,
ionic or of dispersion type, are not revealed. Nonetheless, the coarse treatment
of bond lengths is justified by the cluster separation in Fig.~\ref{fig:clustering}.
We will show that the scoring scheme allows for identification of interesting materials, whose
properties can then be investigated experimentally or using electronic
structure methods.

The scaled bonding criterion described in Equation~\ref{eq:bond} is the same one
employed by Ashton \emph{et al.}~\cite{ashton2017topology} in their study
of layered materials.
An additive bonding criterion of the form
$d_{ij} < r_i + r_j + \Delta$ is used by 
Mounet \emph{et  al.}\cite{Mounet:2018ks} and Cheon \emph{et al.}~\cite{cheon2017weaklybonded},
using van-der-Waals radii and elemental radii respectively.
In these works, the material dimension is determined by sampling a range of
parameter values (either $k$ or $\Delta$) in a fixed interval, which does not
easily permit the construction of a scoring parameter.  Cluster dimensionalities are determined using a
topology-scaling algorithm (TSA)~\cite{ashton2017topology} (also proposed
in \cite{cheon2017weaklybonded}), which relates the dimension
to the number of bonded clusters as a function of
the size of a periodic supercell, or using the RDA~\cite{Mounet:2018ks}.
Due to the use of a fixed-size supercell, the TSA and RDA can respectively underestimate and overestimate
the number of bonded clusters in certain materials with complex geometries.
In the methods section we describe a variant of the RDA which correctly assigns all atoms to bonded clusters
without the need to specify a supercell.
Except for such complicated cases, however, our definition agrees with the TSA
and the supercell RDA.

Other methods for identification of layered materials include the
analysis of the packing fraction~\cite{bjorkman2012vdwbonding, lebegue2013twodimensional},
identification of layered slab structures~\cite{gorai2016slab},
and the use of discrepancies between experimental lattice constants and those predicted by
density functional theory (DFT)~\cite{choudhary2017lattice}.
By identifying structures with interlayer sodium atoms, Zhang~\emph{et al.}~\cite{zhang2018sib} have
investigated promising layered cathode materials for sodium-ion batteries.
McKinney~\emph{et al.}~\cite{mckinney2018ionic} have extended this search to general `ionic layered' structures.


\begin{table}
\begin{tabular}{|r|r|r|r|r|}
\hline
		& \multicolumn{2}{c|}{COD}	& \multicolumn{2}{c|}{ICSD}\\
\cline{2-5}
					& Removed & Remain.	& Removed & Remain.\\
\hline
Initial					&      - & 400731       &     - & 184754\\
$> 200$ atoms			& 185329 & 215402       &  7474 & 177280\\
Partial occupancy		&  49015 & 166387       & 75659 & 101621\\
Missing hydrogen		&   1470 & 164917       &  7184 &  94437\\
Defective				&  10219 & 154698       &  5703 &  88734\\
Duplicates				&  15646 & 139052       & 60019 &  28715\\
\hline
Total remaining & \multicolumn{4}{c|}{167767}\\
\hline
\end{tabular}
\caption{Number of structures remaining after each stage of filtering, performed in the 
order shown.  `Defective' structures encompass incorrect CIF files and theoretical structures, and manually identified entries such as misfit compounds, surface structures, and superstructures.
Where duplicate structures are found across the two databases, the COD structure is kept.
\label{table:filtering}}
\end{table}


\begin{table}
\centering
\begin{tabular}{|r||r|r|r|r|}
\hline
Dim.  &       0 &     1 &    2 &     3 \\
\hline
    0 &  105199 &       &      &       \\
    1 &    3503 &  3285 &      &       \\
    2 &    2946 &    15 & 4623 &       \\
    3 &    3010 &    22 &    0 & 45148 \\
\hline
\end{tabular}
\caption{Number of entries of each dimensionality type found in the
  ICSD and COD. In the diagonal the number of materials with a single
  dimension are shown while the off-diagonal entries indicate
  materials with components of two different dimensionalities. In
  addition to the single and bi-dimensional materials counted here, we
  have found 16 tri-dimensional structures with 0D, 1D, and 2D components.
\label{table:abundance}}
\end{table}

{
\setlength{\tabcolsep}{0pt} 
\setlength{\fboxsep}{0pt}%
\setlength{\fboxrule}{1pt}%

\def\hrulefill{\leavevmode\leaders\hrule height 2pt\hfill\kern0pt}
\newcommand*\ruleline[1]{\par\noindent\raisebox{.8ex}{\makebox[0.96\linewidth]{\hrulefill\hspace{1ex}\raisebox{-.8ex}{#1}\hspace{1ex}\hrulefill}}}

\newcommand*\partialline[2]{\par\noindent\raisebox{.8ex}{\makebox[#1]{\hrulefill\hspace{1ex}\raisebox{-.8ex}{#2}\hspace{1ex}\hrulefill}}}

\begin{figure*}
\centering
\begin{tabular}{
C{0.199\textwidth}
C{0.199\textwidth}
C{0.199\textwidth}
C{0.199\textwidth}
C{0.199\textwidth}}
\multicolumn{5}{c}{\ruleline{\textbf{1D}}}\\
\footnotesize{COD 4344111}
& \footnotesize{ICSD 79796}
& \footnotesize{ICSD 428184}
& \footnotesize{ICSD 33693}
& \footnotesize{ICSD 238683}
\\
\fbox{\includegraphics[width=0.164\textwidth]{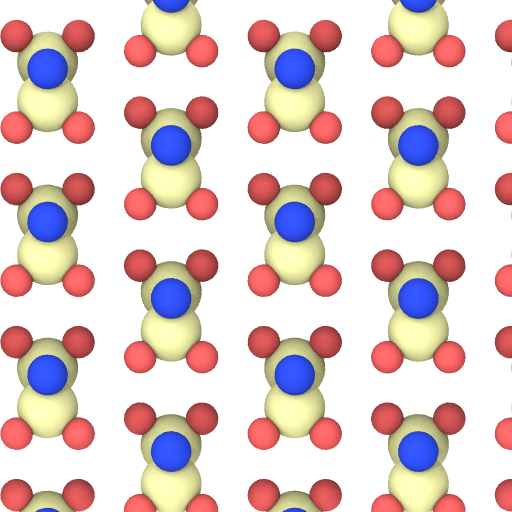}}
&\fbox{\includegraphics[width=0.164\textwidth]{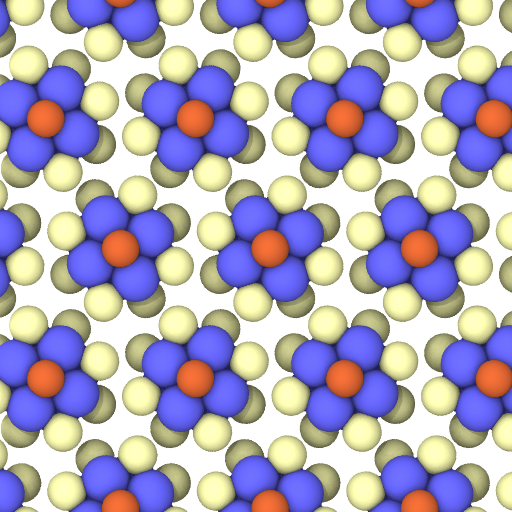}}
&\fbox{\includegraphics[width=0.164\textwidth]{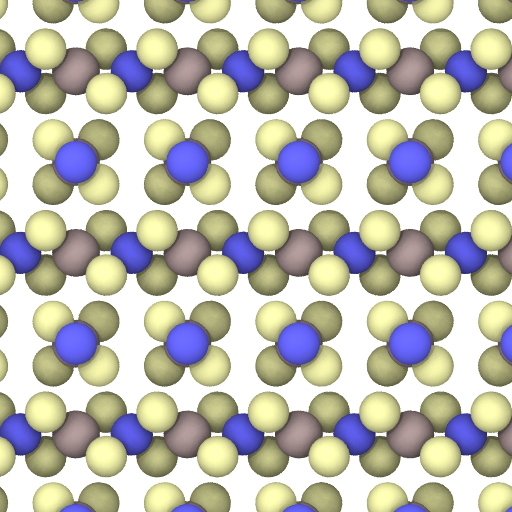}}
&\fbox{\includegraphics[width=0.164\textwidth]{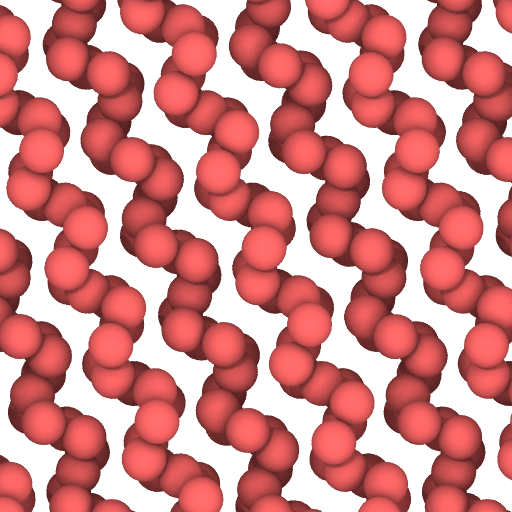}}
&\fbox{\includegraphics[width=0.164\textwidth]{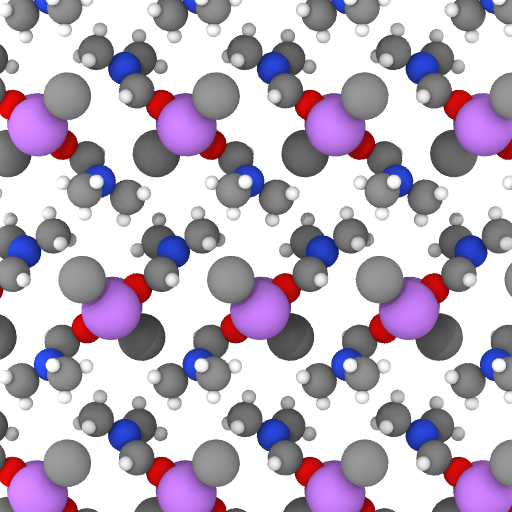}}
\\
\ch{N P F2}
&\ch{Si Ta4 Te4}
&\ch{Al (P S4)}
&\ch{S}
&\ch{Li Cl C3 H7 N O}
\\
$s_{1\text{D}}=0.984$
& $s_{1\text{D}}=0.863$
& $s_{1\text{D}}=0.958$
& $s_{1\text{D}}=0.940$
& $s_{1\text{D}}=0.966$
\\\\
\multicolumn{3}{c}{\partialline{0.56\linewidth}{\textbf{2D}}}
& \multicolumn{2}{c}{\partialline{0.37\linewidth}{\textbf{3D}}}\\
\footnotesize{COD 9007661}
& \footnotesize{ICSD 163023}
& \footnotesize{ICSD 188831}
& \footnotesize{ICSD 411179}
& \footnotesize{ICSD 164652}
\\
\fbox{\includegraphics[width=0.164\textwidth]{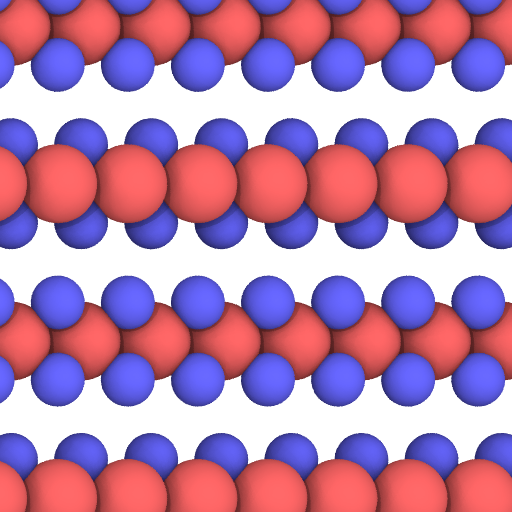}}
&\fbox{\includegraphics[width=0.164\textwidth]{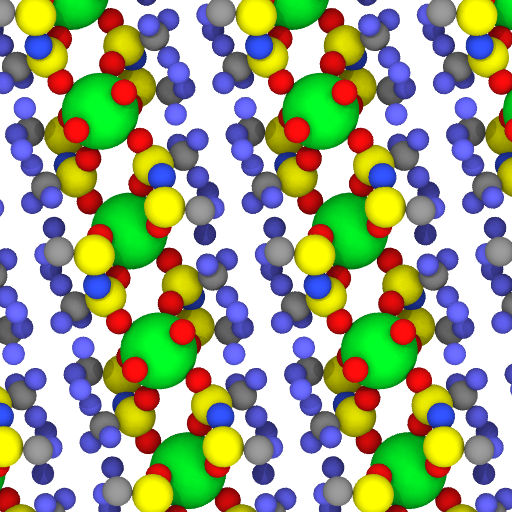}}
&\fbox{\includegraphics[width=0.164\textwidth]{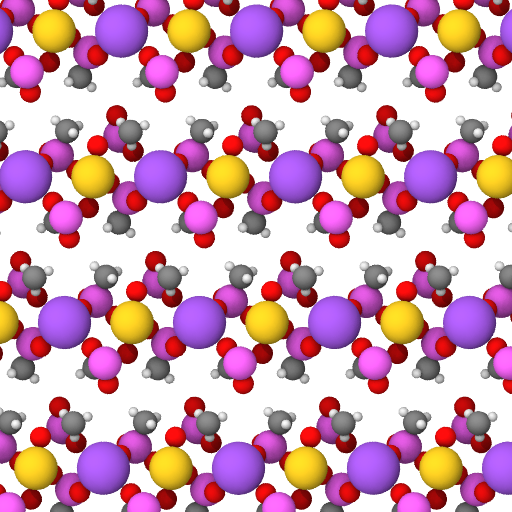}}
&\fbox{\includegraphics[width=0.164\textwidth]{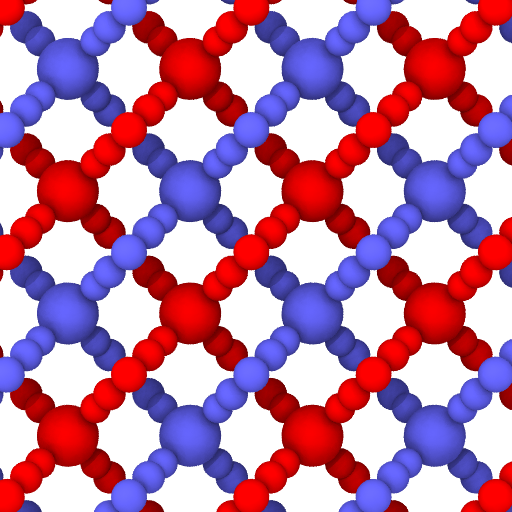}}
&\fbox{\includegraphics[width=0.164\textwidth]{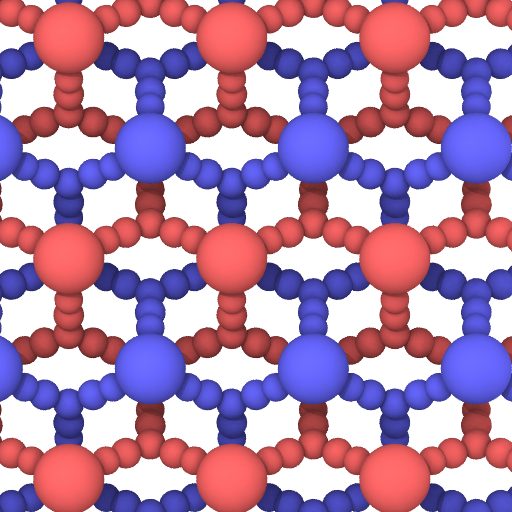}}
\\
\ch{Mo S2}
&\ch{Sr ((C F3 S O2)2 N)2}
&\ch{Na (Au (C H3 S O3)4)}
&\ch{Ag (B (C N)4)}
&\ch{Ca (C (C N)3)2}
\\$s_{2\text{D}}=0.959$
& $s_{2\text{D}}=0.981$
& $s_{2\text{D}}=0.974$
& $s_{3\text{D}}=0.936$
& $s_{3\text{D}}=1.000$
\\\\
\multicolumn{2}{c}{\partialline{0.37\linewidth}{\textbf{1D+2D}}}
& \multicolumn{3}{c}{\partialline{0.56\linewidth}{\textbf{1D+3D}}}\\
\footnotesize{COD 7222569}
& \footnotesize{ICSD 248420}
& \footnotesize{COD 1529916}
& \footnotesize{COD 9000964}
& \footnotesize{ICSD 412254}
\\
\fbox{\includegraphics[width=0.164\textwidth]{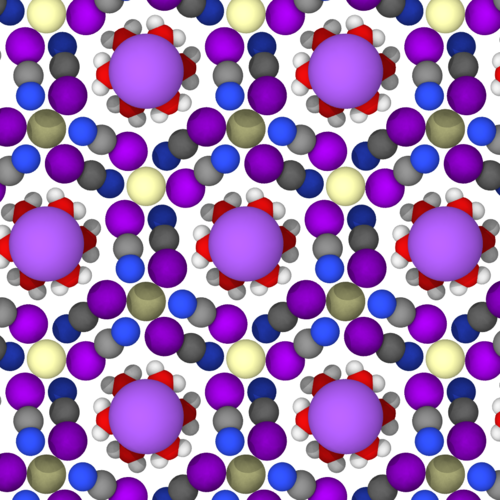}}
&\fbox{\includegraphics[width=0.164\textwidth]{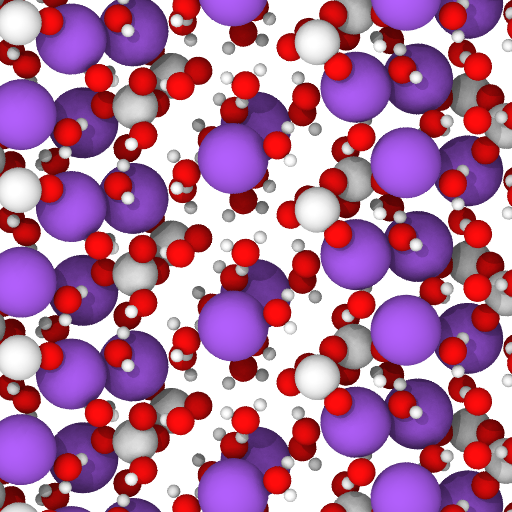}}
& \fbox{\includegraphics[width=0.164\textwidth]{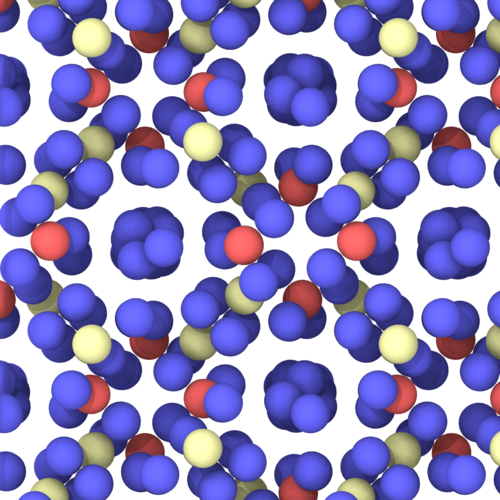}}
&\fbox{\includegraphics[width=0.164\textwidth]{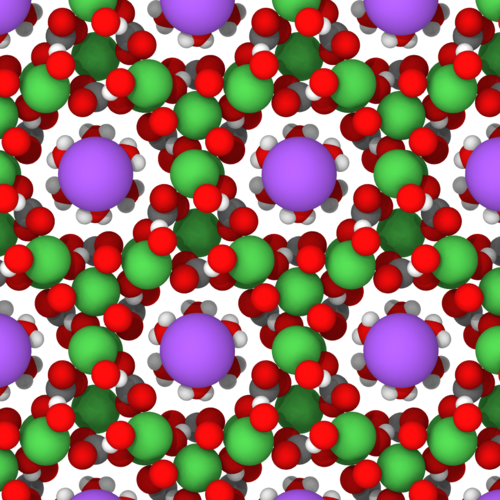}}
&\fbox{\includegraphics[width=0.164\textwidth]{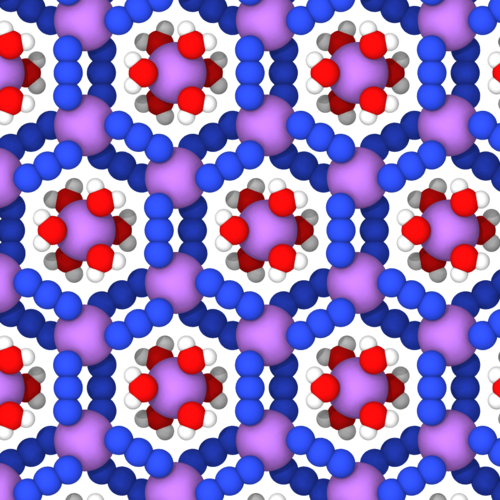}}
\\
\ch{Na (H2 O)3}
&\ch{Na3 H P2 O7}
&\ch{S29 Ta4 P4}
&\ch{Na2 (H2 O)6}
&\ch{Li (N3) (H2 O)}
\\
\ch{(Mn (N C S)3)}
&\ch{(H2 O)9}
&\ch{}
&\ch{Ni8 (C O3)6 (O H)6}
&\ch{}
\\
$s_{1\text{D}+2\text{D}}=0.550$
& $s_{1\text{D}+2\text{D}}=0.815$
& $s_{1\text{D}+3\text{D}}=0.936$
& $s_{1\text{D}+3\text{D}}=0.742$
& $s_{1\text{D}+3\text{D}}=0.595$
\end{tabular}
\caption{Smorgasbord of low-dimensional and mixed-dimensional
  materials, identified by applying the automatic dimensionality
  classification method to materials in the
  ICSD.
  \label{fig:smorgasbord}}
\end{figure*}
}

\subsection{Mining the ICSD and COD}

We have analyzed all materials in the ICSD and COD using the proposed scoring
parameter. Fig.~\ref{fig:smorgasbord}
shows examples of materials with different dimensionalities and high
values of the scoring parameter.

The database has been filtered in standard ways~\cite{kirklin2015oqmd, Mounet:2018ks} by removing
incomplete and/or defective entries, structures with more than 200 atoms, structures with partial occupancies, theoretically calculated structures, and structures with missing hydrogen atoms.
Duplicate structures are removed using the structure matcher function of pymatgen~\cite{ong2013pymatgen}.
The filtering process reduces the initial set of 585485 CIF files to 167767 structures.  The filtering statistics are shown in Table~\ref{table:filtering}.

An overview of the database is shown in
Table~\ref{table:abundance}. In this table the materials have all been
categorized by the dimension(s) with the largest value of the scoring
parameter $s$. In some cases all $s$ values may be fairly small and the
classification is then rather uncertain.  A large number of materials (105199) are classified as 0D. These
are mostly molecular crystals, which we shall not consider any further here.

The second largest category is the 3D materials. Most of these have a
single 3D component, but some of them have two components still with
large $s$-values. Two examples [Ag(B(CN)$_4$ and Ca(C(CN)$_3$)$_2$]
are shown in Fig.~\ref{fig:smorgasbord}. As can be seen from the
figure, both materials consist of two identical interpenetrating
networks which cannot be disentangled without breaking bonds.  The two
networks are sufficiently spatially separated to give scoring values
above 0.7. (In the figure the two networks are colored red and
blue). 

4623 materials are identified as two-dimensional, which is about 2.8\%
of all materials. This can be compared to for example the study by
Mounet \emph{et al.}~\cite{Mounet:2018ks} where they find 1825 out of
108423 materials (or also about 1.7\%) of the materials to be easily or
potentially exfoliable.  About 2\% of the materials are classified as 1D.

There are also some materials with several components of different
dimensionality. In particular there are 9459 materials which have one
or more 0D components in combination with components of higher
dimensionality. These correspond to molecules or molecular ions
embedded in the higher dimensional network. Only a few materials combine
1D, 2D, and 3D components. We find 15 materials combining 1D and
2D. There are 22 materials which combine 1D and 3D components. Three
of them are shown in Fig.~\ref{fig:smorgasbord}.

While we have made every effort to remove inconsistent structures
from the database, an automated filtering is not sufficient given the
many different types of errors and partial structures present in the
ICSD and COD.  The numbers presented here should therefore be taken as only
approximate.

A database containing the calculated scoring parameters for all
dimensionalities for all compounds in the ICSD and COD is available
at the Computational Materials Repository\cite{cmr_database}.

\begin{figure*}
\centering
\includegraphics[width=0.85\textwidth]{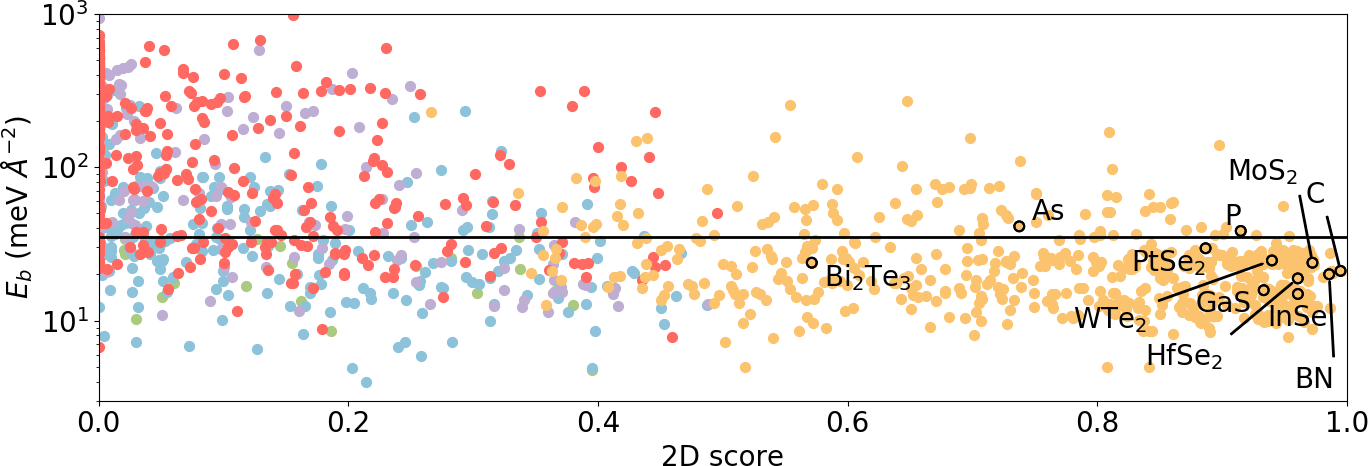}\vspace{2mm}\\
\textbf{Classification:}\hspace{3mm}
\includegraphics[width=0.06\columnwidth]{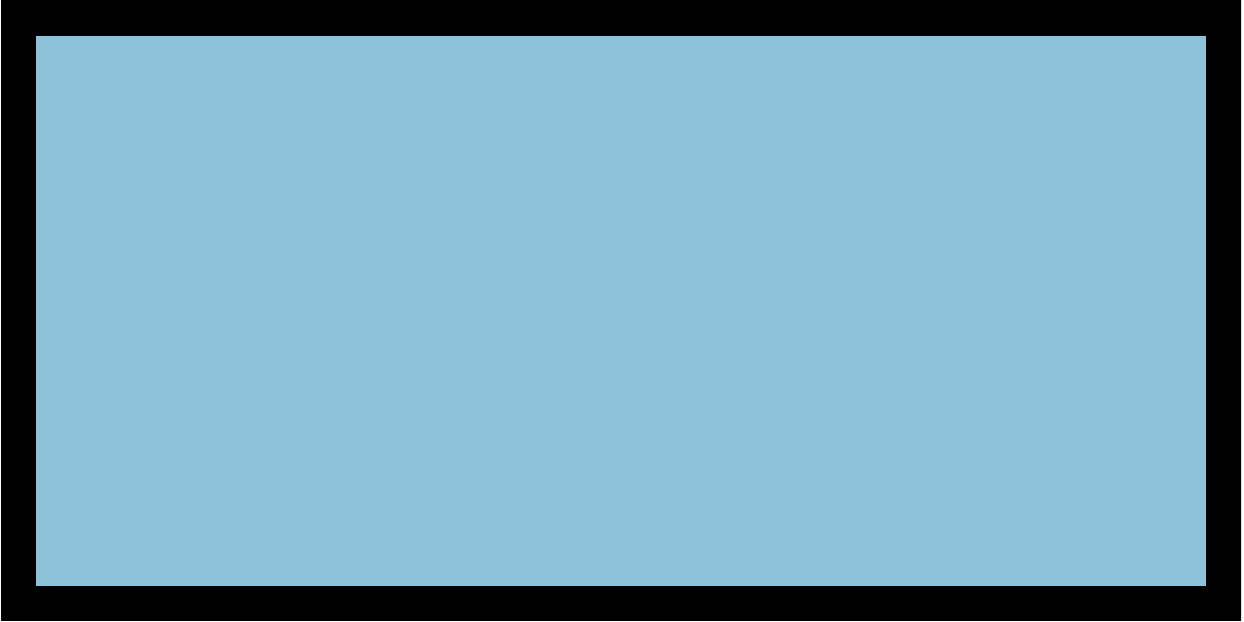} \textbf{0D}\hspace{3mm}
\includegraphics[width=0.06\columnwidth]{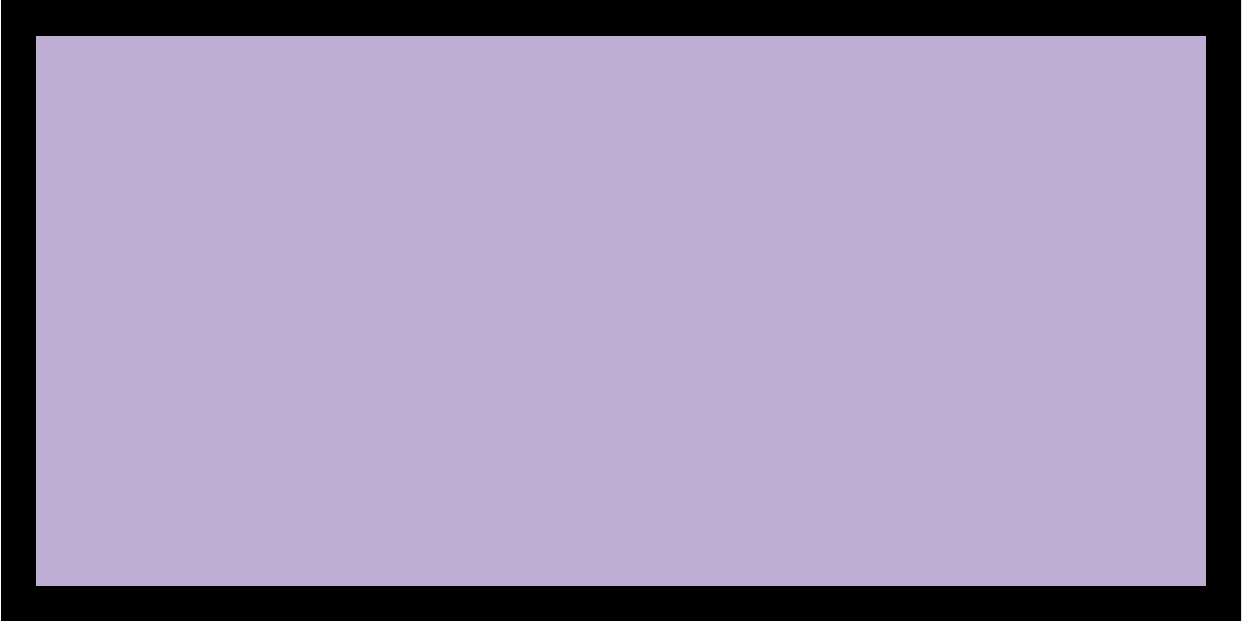} \textbf{02D}\hspace{3mm}
\includegraphics[width=0.06\columnwidth]{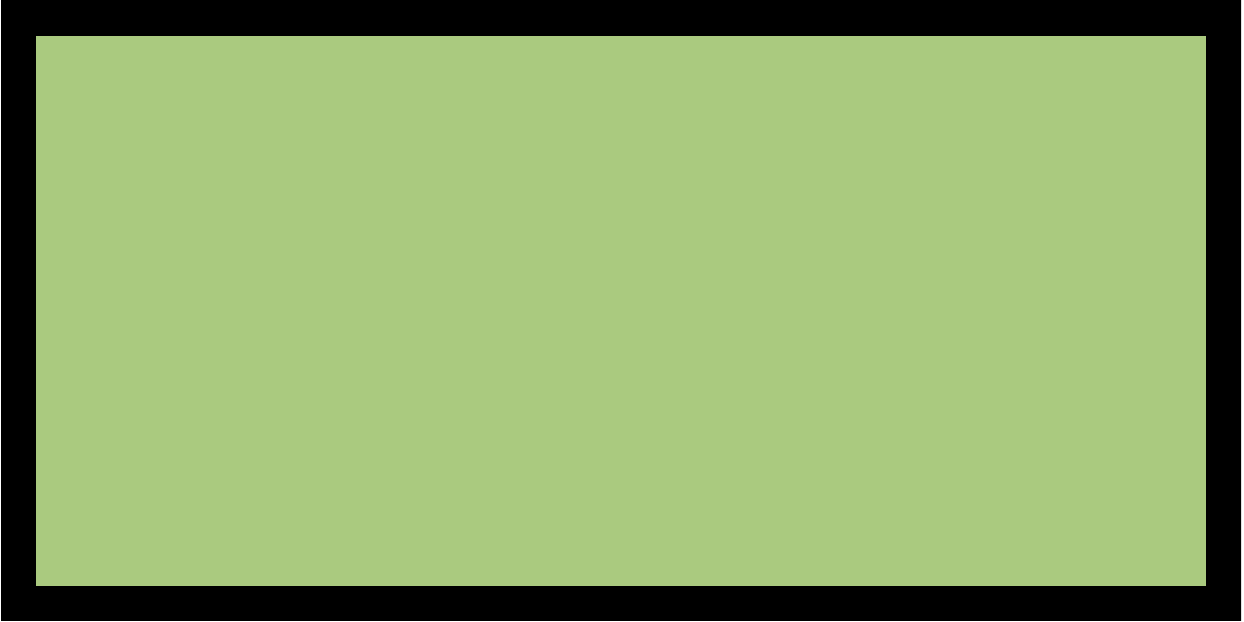} \textbf{1D}\hspace{3mm}
\includegraphics[width=0.06\columnwidth]{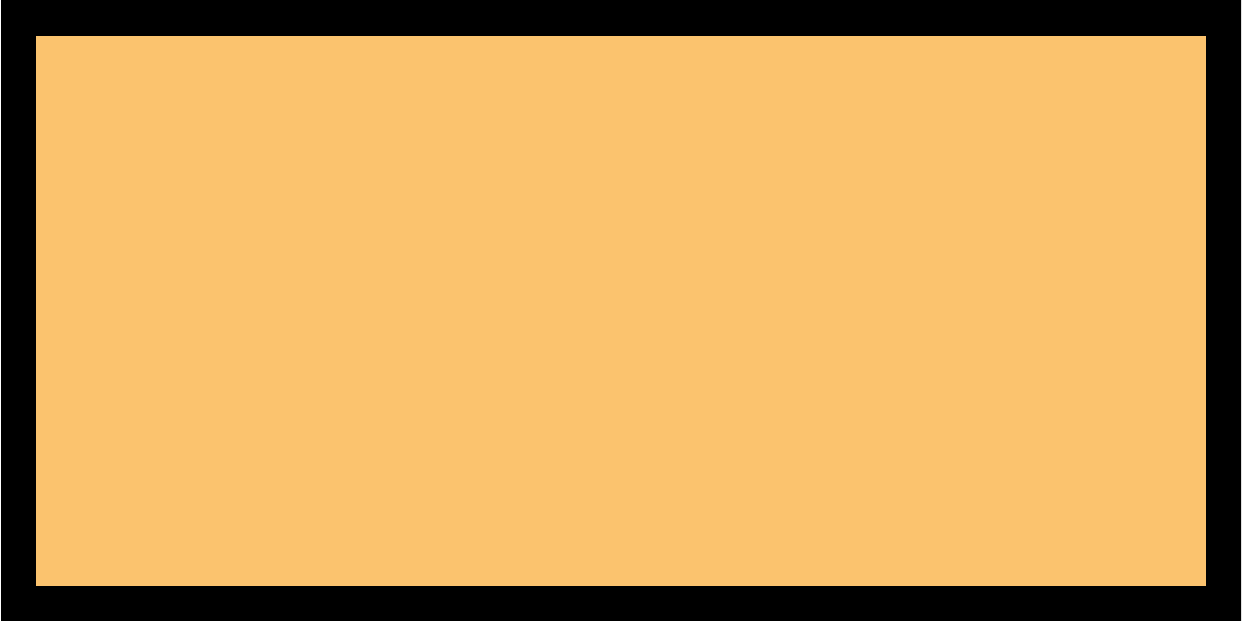} \textbf{2D}\hspace{3mm}
\includegraphics[width=0.06\columnwidth]{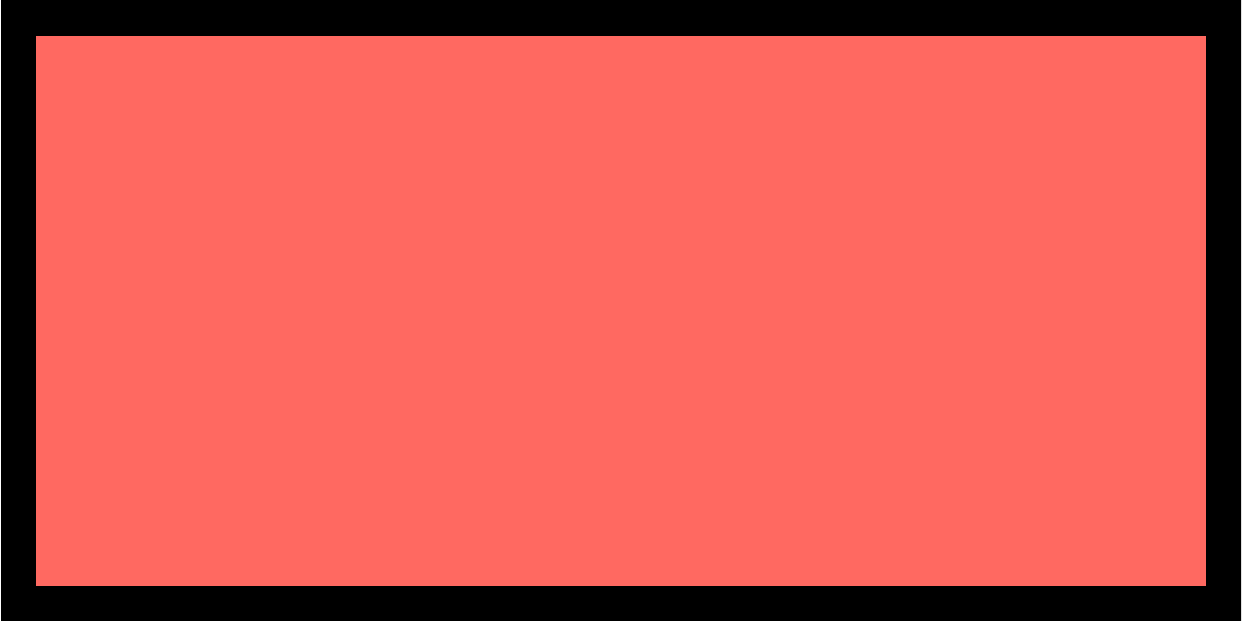} \textbf{3D}
\caption{ Binding energies ($E_b$) vs.~2D scores for 1535 layered
  materials identified by Mounet \emph{et al.}~\cite{Mounet:2018ks},
  colored according to dimensionality classification.  Structures with $E_b < 35 \text{mev} / \text{\AA}^2$
  are classified by Mounet \emph{et al.}~as \emph{easily exfoliable}.
  For clarity, 44 01D, five 03D, and four 012D structures are not shown here.
}
\label{fig:marzari_comparison}
\end{figure*}

\begin{table}
\centering
{
\begin{tabular}{|c|c|c|c|c|c|}
\hline
Source & ID & Compound & $s$ & $k_1$ & $k_2$\Bstrut{2}\\
\hline
COD & 1000410 & \ch{TlAlF4} & 0.987 & 0.992 & 2.330\Bstrut{1.2}\\
COD & 9000046 & \ch{C} & 0.986 & 0.933 & 2.251\Bstrut{1.2}\\
ICSD & 27987 & \ch{BN} & 0.984 & 0.935 & 2.170\Bstrut{1.2}\\
ICSD & 248325 & \ch{C3N4} & 0.983 & 0.910 & 2.155\Bstrut{1.2}\\
ICSD & 187384 & \ch{Rb (Au (C F3 S O3)4)} & 0.982 & 1.010 & 2.313\Bstrut{1.2}\\
ICSD & 163023 & \ch{Sr((CF3SO2)2N)2} & 0.981 & 1.014 & 2.516\Bstrut{1.2}\\
COD&  1525422 & \ch{K3 Mn (C N)6} & 0.980 & 0.935 & 2.039\Bstrut{1.2}\\
COD & 1534338 & \ch{MgCl2} & 0.977 & 0.959 & 1.976\Bstrut{1.2}\\
ICSD & 161278 & \ch{B3C10N3} & 0.977 & 0.965 & 1.968\Bstrut{1.2}\\
COD & 2242431 & \ch{Cs(N(SO2CF3)2)} & 0.977 & 0.965 & 1.968\Bstrut{1.2}\\
\hline
\end{tabular}
}
\caption{Top ten `most 2D' materials in the ICSD and COD, as ordered by the interval scoring method in Equation~(\ref{eq:scoring_scheme_s}).  Since the databases contain many layered polymers with large
interlayer spacings, we have not included structures containing hydrogen in this list.
\label{table:most_2d}}
\end{table}

\begin{table}
\centering
{
\begin{tabular}{|c|c|c|c|c|c|}
\hline
Source & ID & Compound & $s$ & $k_1$ & $k_2$\Bstrut{2}\\
\hline
COD & 4344111 & \ch{NPF2} & 0.984 & 0.889 & 2.184 \Bstrut{1.2}\\
COD & 9010982 & \ch{SO3} & 0.983 & 0.954 & 2.133 \Bstrut{1.2}\\
ICSD & 72577 & \ch{Ru(CO)4} & 0.982 & 0.979 & 2.095 \Bstrut{1.2}\\
ICSD & 47183 & \ch{Cu Co (CO)4} & 0.980 & 0.926 & 2.044 \Bstrut{1.2}\\
ICSD & 291211 & \ch{SiS2} & 0.967 & 0.986 & 1.815 \Bstrut{1.2}\\
ICSD & 22090 & \ch{RuCl3} & 0.967 & 0.940 & 1.812 \Bstrut{1.2}\\
ICSD & 415951 & \ch{V (Al Cl4)2} & 0.965 & 0.990 & 1.784 \Bstrut{1.2}\\
ICSD & 78778 & \ch{CrF4} & 0.964 & 0.985 & 1.780 \Bstrut{1.2}\\
ICSD & 428185 & \ch{Al P S4} & 0.964 & 0.995 & 1.771 \Bstrut{1.2}\\
ICSD & 419661 & \ch{CrF5} & 0.962 & 0.996 & 1.751 \Bstrut{1.2}\\
\hline
\end{tabular}
}
\caption{Top ten `most 1D' hydrogen-free materials in the ICSD and COD, as ordered by the
interval scoring method in Equation~(\ref{eq:scoring_scheme_s}).
\label{table:most_1d}}
\end{table}

\subsection{Physical Significance of the Scoring Parameter}

Due to the well-defined identification of the 2D materials, the
scoring scheme also serves as a simple predictor of exfoliability.
Mounet \emph{et al.}~\cite{Mounet:2018ks} have calculated the
exfoliation energy ({\it i.e.} the binding energy between layers) of
1535 layered materials, and they suggest an energy of
35 meV/\AA$^{2}$ as the threshold for ``easily
exfoliable'' materials. They furthermore highlight 11 materials,
which they denote as ``well-known'' 2D materials.

In Fig.~\ref{fig:marzari_comparison} we show the calculated exfoliation
energies versus the scoring parameter $s_2$. There is a clear
correlation between the scoring parameter and the exfoliation energy
with essentially all of the high-scoring materials (say $s_2>0.7$)
having an exfoliation energy below the threshold. The separation of
materials of different dimensionality is also clearly seen here by the low
density of points in the region $s_2 \approx 0.3-0.5$. The 11
well-known 2D materials are also shown in the figure. All of them,
except \ch{Bi2Te3}, have high scoring values with $s_2>0.7$.  Despite
its small interlayer distance, \ch{Bi2Te3} is nonetheless classified as
a 2D material, since $s_2$ is larger than its other scores.

It should be noted that although the exfoliation energy is a highly
relevant quantity for the exfoliation process, it is not clear whether
an absolute threshold in energy is the best indicator of
exfoliability. The exfoliation process involves breaking the bonds
between the layers keeping the bonds within the layers intact, so the
exfoliation energy should be seen relative to the intralayer bond
strengths. While the scoring parameter proposed here does not
explicitly involve the energetics, the high-scoring materials have a
clear separation between the intra- and intercomponent bond lengths,
which can be expected to be a characteristic of easily exfoliable materials.

\subsection{Ranking of Low-dimensional Materials}

In addition to dimensionality classification, the scoring parameter
defines an order on materials.  We have identified the ten materials
in the ICSD and COD with the highest 2D scores, shown in
Table~\ref{table:most_2d}.
Widely studied layered structures such as graphene, boron nitride, 
and magnesium chloride are highly ranked.  Some of the remaining structures
have much larger unit cells, but are
nonetheless clearly van der Waals bonded layered structures.
It should be noted that the detailed ordering of the top materials is sensitive to the
detailed choice of the function $f(x)$ in
Equation~\ref{eq:scoring_scheme_f}, whereas the overall classification
of the materials is more robust.

Similarly a list of the highest-scoring 1D materials is provided in
Table~\ref{table:most_1d}. We shall not discuss these materials in
depth here, but briefly characterize the top five entries with two or
three different chemical elements. For all of these the
one-dimensional or chain-like character has already been
recognized. \ch{NPF2} consists of chains of alternating nitrogen and
phosphorous atoms with the fluorine atoms bound to the
phosphorous. The chains can also close on themselves forming
ring-shaped molecules. \ch{SO3} is an asbestos-like structure made up
of corner-linked \ch{SO4} tetrahedra forming spiraling chains, while
the chains in the \ch{SiS2} structure consists of edge-sharing
tetrahedra. The two ruthenium compounds form chains of ruthenium
atoms.  \ch{Ru(CO)4} is constructed from planar units with ruthenium
in the middle and CO molecules attached with a fourfold rotation
symmetry. These units are then stacked forming chains of ruthenium.
\ch{RuCl3} is in the $\beta$ phase also called the \ch{ZrCl3}
structure. Here again the ruthenium atoms form linear chains, but with
the chlorine atoms connecting two adjacent ruthenium atoms. It can be
noted that \ch{RuCl3} also appears as a strongly layered material
($s_2$ = 0.933) in the $\alpha$ phase with prototype \ch{RhBr3}. We
have performed density functional calculations for these highly 1D
compounds using the GPAW code \cite{Mortensen:2005ep,
  Enkovaara:2010jd} and the Atomic Simulation Environment (ASE)
\cite{Bahn:2002to, Larsen:2017hn}. 
The three compounds \ch{NPF2}, \ch{SO3}, and \ch{SiS2}
are found to be non-magnetic large band gap semiconductors. 

In the two compounds with ruthenium chains the
distance between the ruthenium atoms are in fact comparable to the
bond distance in ruthenium bulk metal. However, the strong couplings
to the attached atoms and molecules lead to opening of band gaps.
According to the DFT calculations \ch{Ru(CO)4} is non-magnetic while
\ch{RuCl3} is found to be antiferromagnetic. Details of the calculations 
can be found in the Supplementary Material.

The scoring approach also allows for identification of materials with
several components of different dimensionality. Five of these are
shown in Fig.~\ref{fig:smorgasbord}. The two 1D+2D materials and the
last two 1D+3D materials all involve alkali atoms (Na or Li) decorated
with water molecules as the 1D component. Note that in the case of
\ch{Na (H2 O)3 (Mn (N C S)3} the chains penetrate the 2D layers in the
framework, while in the case of \ch{Na3 H P2 O7 (H2 O)9} the chains
run parallel to the 2D components.

In these materials charge transfer takes place with approximately one
electron per alkali atom donated to the 2D or 3D framework. We have
investigated this by performing DFT calculations followed by a Bader
analysis, \cite{Bader:1990we,Tang:2009cb} where an electronic charge
is associated with each atom based on a natural per-atom division of
the electronic density. For the compounds \ch{Na (H2 O)3 (Mn (N C S)3}
and \ch{Na2 (H2 O)6 Ni8 (C O3)6 (OH)6} the chains consist of \ch{Na
  (H2 O)3} units with an electron transfer of 0.85 and 0.90 electrons
per unit, respectively. Similarly in the case of \ch{Li (N3)(H2O)},
the charge transfer is 0.85 per \ch{Li (H2 O)3} unit. \ch{Na3 H P2 O7
  (H2 O)9} contains chains of \ch{Na (H2 O)4} molecular ions with a
charge transfer of 0.80 electrons per unit.

The charge transfer in these systems illustrates that the geometrically
defined scoring parameter does not only identify components which are
exclusively van der Waals bonded to each other. Charge transfer may
take place between spatially separated components giving rise to
bonding of a more ionic character.

\begin{figure}
\includegraphics[width=0.95\columnwidth]{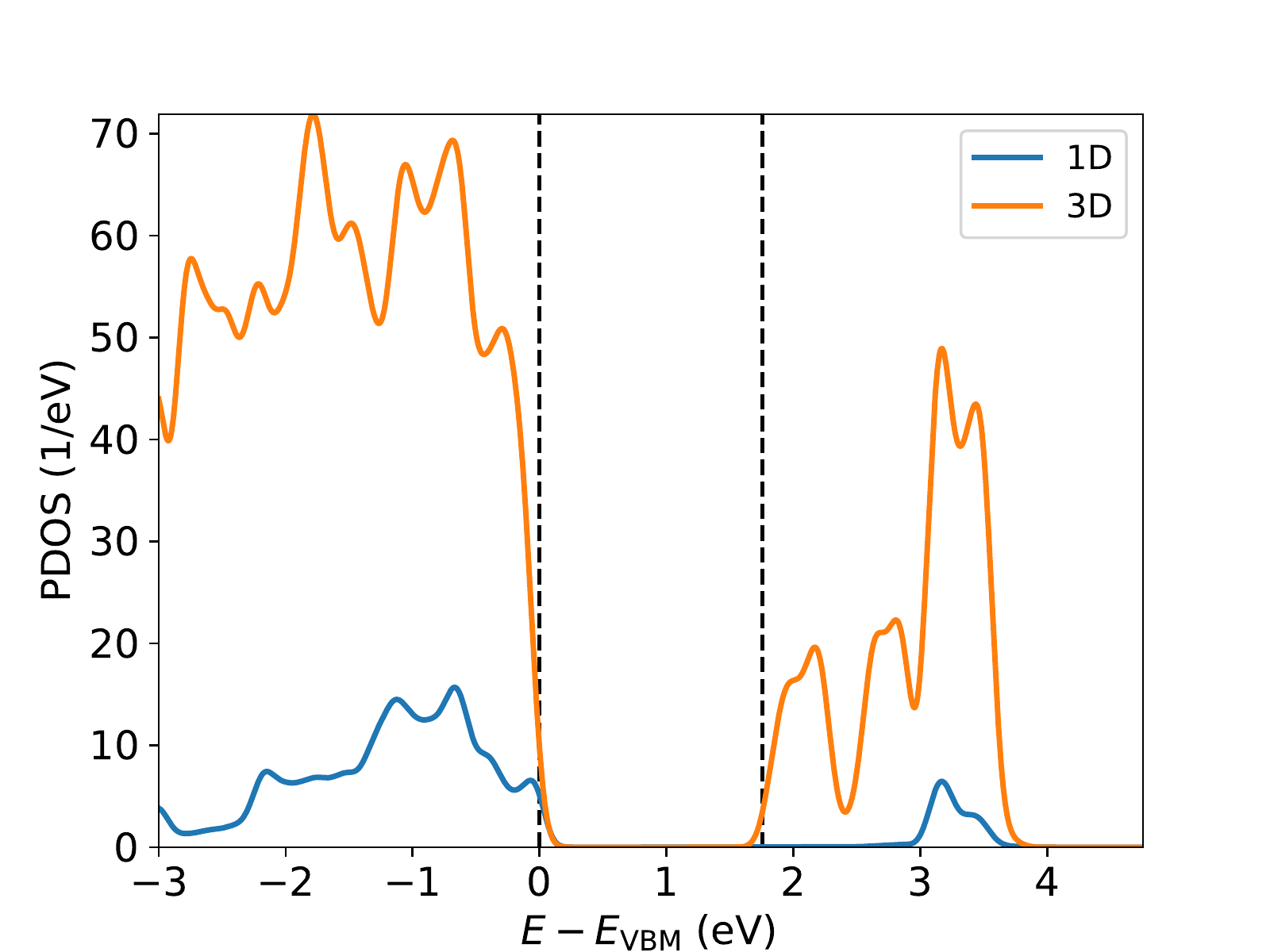}
\caption{Density of states projected onto the 1D and 3D components of
the \ch{S29 Ta4 P4} compound. The energy is relative to the valence
band maximum (VBM) and the vertical dashed lines indicate the band edges
of the smallest band gap. The electronic spectrum is calculated using
    the GLLB-SC exchange-correlation functional.}
\label{fig:dos}
\end{figure}

The last compound in Fig.~\ref{fig:smorgasbord} of mixed
dimensionality is \ch{S29 Ta4 P4}. It has a very intriguing structure.
It consists of spiraling sulfur
chains penetrating a 3D network constructed of \ch{Ta P S6} building
blocks. The 3D network itself consists of two identical interpenetrating 
components.
The ability of Ta-P-S compounds to form tunnels has previously
been reported \cite{Evain:1987do} and sulfur spirals appear in
several compounds. The present compound does not exhibit any charge
transfer between the components. Fig.~\ref{fig:dos} shows the
calculated density of states for the \ch{S29 Ta4 P4} compound
projected onto the 1D and 3D components. The two components are seen
to exhibit different band gaps. This opens the possibility for
selectively exciting electrons in one of the components using light
with an appropriate wavelength.

\section{Conclusion}
We have defined a simple geometric scoring parameter to
identify materials of particular dimensionality. The parameter
provides an estimate of the degree to which a given dimensionality is
present in the compound. The parameter is easy to calculate and can be
applied to large materials databases. As mentioned in the introduction
several computational 2D materials databases are presently under
construction while 1D materials and materials of mixed dimensionality
have received much less attention. The present approach allows for
simple identification of existing 1D or mixed-dimensional materials,
which can form templates that can be used to construct larger
computational databases for materials of a given dimensionality.

\section{Methods}

\subsection{Component Dimensionality}
\label{sec:dimensionality_problem}

\begin{figure}
  \centering
  \includegraphics[width=0.8\columnwidth]{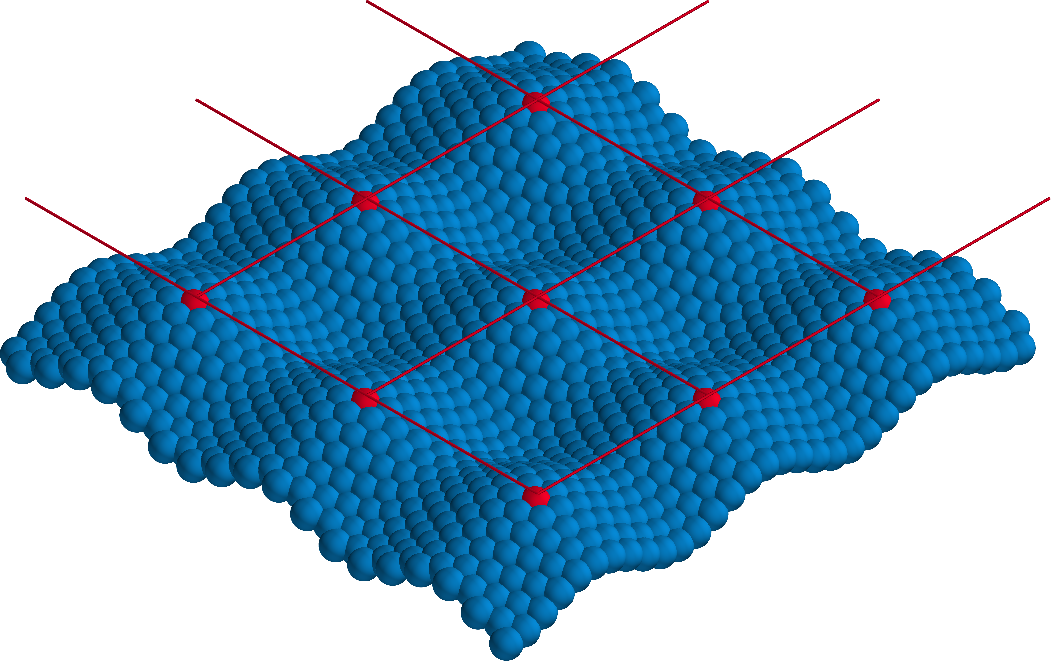}
  \caption{Cut-out of a periodic corrugated 2D component.  The
    component dimensionality can be found by selecting any atom in the
    component, and identifying all other atoms in the bonded cluster
    with the same fractional coordinates, shown here in red.  The rank
    of the subspace spanned by these atoms (here, a plane) determines
    the dimensionality.}
  \label{fig:definition}
\end{figure}

A material will in general consist of several clusters of bonded
atoms. Such clusters we term the \emph{components} of the
material. The components may have different dimensionalities and they
should therefore be investigated separately.

Our definition of material dimensionality of a component is as
follows: select an atom in the component, with atomic coordinates
$\set{x}_1$.  Let
$\set{X} = \{ \set{x}_1, \set{x}_2, \set{x}_3, \ldots, \set{x}_i \}$
denote the set of atoms to which the first atom is bonded, and which
have the same fractional coordinates but in different unit cells,
i.e., $\set{x}_i = \set{x}_1 + \set{C}^T \set{h}_i$, where \set{C} is
the unit cell description and $\set{h}_i$ is an integer vector. Then,
the component dimensionality is the rank of the subspace spanned by
\set{X}:

\begin{equation}
  \text{dim}\left(\set{X}\right) = \text{rank}\left( \{\set{x}_2 - \set{x}_1, \set{x}_3 - \set{x}_1, \ldots, \set{x}_i - \set{x}_1 \} \right)
\end{equation}
This definition (illustrated in Fig.~\ref{fig:definition})
accommodates both corrugation and thickness. While \set{X} is an
infinite set for all but 0D components,
$\text{dim}\left(\set{X}\right)$ can be determined in a finite number
of steps by exploiting the periodicity of a material.

As described above, determination of the dimension of a
material requires an analysis of its constituent bonded clusters, or
\emph{components}.  To find the dimension of a component,
the rank determination algorithm
(RDA) of Mounet~\emph{et al.}~\cite{Mounet:2018ks} uses a supercell of fixed
size with open boundary conditions.  If the supercell is too small, the number of components
might be overestimated.

Conversely, the topological scaling algorithm (TSA) of Ashton~\emph{et
  al.}~\cite{ashton2017topology} uses periodic unit
cells, which can underestimate the number of components by forming improper
connections between them.  By improper connections, we mean components which are disconnected in the infinite
crystal but are connected due to the periodic cell chosen.

\begin{figure*}
\begin{tabular}{cccccccc}
$m=1$ & $m=2$ & $m=3$ & $m=4$ & $m=5$ & $m=6$ & $m=7$ & $m=8$\\
$n_c = 1$ & $n_c = 2$ & $n_c = 1$ & $n_c = 4$ & $n_c = 1$ & $n_c = 2$ & $n_c = 1$ & $n_c = 8$\\
\hspace{1mm}\adjincludegraphics[width=0.095\textwidth, valign=T]{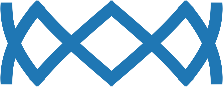}
\hspace{1mm} & \hspace{1mm}\adjincludegraphics[width=0.095\textwidth, valign=T]{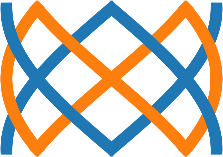}
\hspace{1mm} & \hspace{1mm}\adjincludegraphics[width=0.095\textwidth, valign=T]{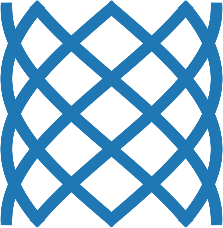}
\hspace{1mm} & \hspace{1mm}\adjincludegraphics[width=0.095\textwidth, valign=T]{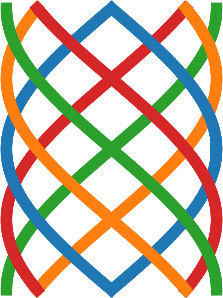}
\hspace{1mm} & \hspace{1mm}\adjincludegraphics[width=0.095\textwidth, valign=T]{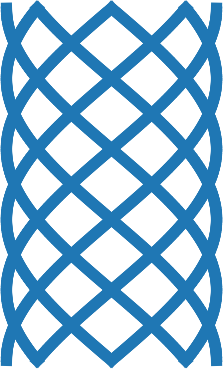}
\hspace{1mm} & \hspace{1mm}\adjincludegraphics[width=0.095\textwidth, valign=T]{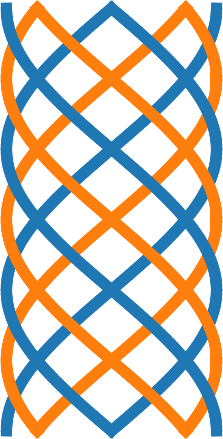}
\hspace{1mm} & \hspace{1mm}\adjincludegraphics[width=0.095\textwidth, valign=T]{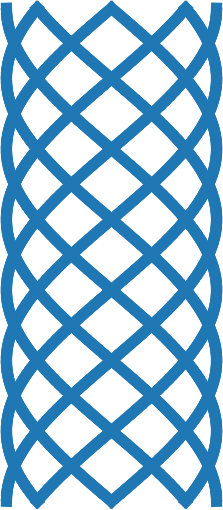}
\hspace{1mm} & \hspace{1mm}\adjincludegraphics[width=0.095\textwidth, valign=T]{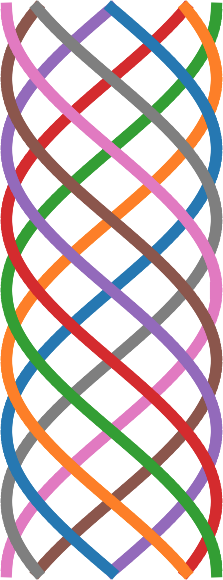}
\hspace{1mm}
\end{tabular}
\caption{Improper connections between components in $n$ helix
  structures, here for $n=8$.  The number of components is denoted by
  $n_c$.  The infinite structure contains 8 components.  Any number of
  repetitions, $m \geq 1$, of the cell for which $m \mod 8 \neq 0$
  results in improper connections between components.}
\label{fig:helices}
\end{figure*}

The problem of improper connections is illustrated with a contrived example in 
Fig.~\ref{fig:helices}, which shows the side view of a selection of
periodic helical structures.  We define an $n$ helix as a structure
which has $n$ components, whose $j^\text{th}$ component has
coordinates given by:
\begin{equation}
x_j = \sin \frac{2 \pi \left(t + j \right)}{n}
\hspace{6mm}
y_j = \cos \frac{2 \pi \left(t + j \right)}{n}
\label{eq:helix_coords}
\end{equation}
The number of components is dependent on the size (along the $t$ axis)
of the periodic cell.  In the formulation given in
Equation~\eqref{eq:helix_coords} any integer is a valid cell length.

Fig.~\ref{fig:helices} shows how the number of components changes
with varying cell periodicity.  In general, the number of components
for an $n$ helix with periodicity $m$ is given by
$\gcd \left( n, m \mod n \right)$.  In order to avoid improper
connections between components, a periodic cell of size $n$ is needed.
This is further complicated for cells containing multiple $n$ helices
of different sizes.  In this case, the correct size of the periodic
cell is given by $\lcm \left( n_1, n_2, n_3, \ldots \right)$, where
$n_i$ denotes the number of components in the $i^\text{th}$ $n$ helix.
For example, a structure containing a $5$-helix, a $6$-helix, and a
$7$-helix requires a periodic cell of length 210.  This cell is so
large that it is unlikely that it would be tested using the existing
methods.

While the example described here is contrived, self-penetrating helical networks have been
assembled experimentally~\cite{xiao2007selfpenetrating,
  yang2012molecularbraids}.  Furthermore, the problem illustrated
has practical consequences: an incorrect periodic cell (such as the
use of a primitive unit cell) causes the interpenetrating polymer
networks shown in Fig.~\ref{fig:smorgasbord} to be misclassified as 2D materials.

\subsection{Algorithm}
\label{sec:dimensionality_algorithm}

Component dimensionalities can be identified using a modified
breadth-first-search (BFS) algorithm, shown in
Algorithm~\ref{alg:graph_search}.  In standard BFS, the search
terminates when all nodes have been visited.  Here, we terminate the
search when the rank of the subspace spanned by a component (i.e.~the
dimensionality) can no longer increase.  The rank of a set of points
is defined as:
\begin{equation}
\text{rank}\left( \set{v} \right) = 
\begin{cases}
    -1 & \text{if } \set{v} = \emptyset\\
    \text{rank}_{M}\left( \set{v} - \vec{v}_1 \right)              & \text{otherwise}
\end{cases}
\end{equation}
where rank$_{M}$ denotes the standard matrix rank.

In this algorithm, components in the aperiodic primitive unit cell are the graph vertices,
and connections between components (across unit cell boundaries) are graph edges.  We note that, by definition,
no edges exist between components within the same cell.

\begin{figure}
\begin{algorithm}[H]
\begin{algorithmic}[1]
\Procedure{CalculateDimensionality}{$\set{E}$, $c$} \label{alg1:input}
	\State $\set{s} := \emptyset$	\label{alg1:s_init}
	\State $\set{v} := \{ \emptyset \hspace{2mm} \forall i \in 1\ldots n \}$	\label{alg1:v_init}
	\State $\set{Q} := \{ \{ c, O \} \}$	\label{alg1:q_init}
	\While {$\set{Q} \neq \emptyset$}	\label{alg1:while_q}

		\State $\{ i, \vec{p} \} := \set{Q}_1$	\label{alg1:q_pop}
		\State $\set{Q} := \set{Q} \setminus \set{Q}_1$	\label{alg1:q_remove}
		\If { $\{ i, \vec{p} \} \in \set{s} $}	\label{alg1:seen_check}
			\State \Continue
		\EndIf

		\State $\set{s} := \set{s} \cup \{ i, \vec{p} \}$	\label{alg1:seen_add}
		\If { $\calcrank{ \set{v}_{i} \cup \{ \vec{p} \} } > \calcrank{ \set{v}_{i} }$ }	\label{alg1:rank_increase}
			\State $\set{v}_{i} := \set{v}_{i} \cup \{ \vec{p} \}$	\label{alg1:v_add}
		\EndIf

		\For { $ \{ j, \vec{\Delta} \} \in \set{E}_{i} $}	\label{alg1:edge_loop}

			\State $\vec{q} := \vec{p} + \vec{\Delta}$	\label{alg1:new_vertex}
			\If {$\{ j, \vec{q} \} \in \set{s} $}	\label{alg1:seen_check_new}
				\State \Continue
			\EndIf

			\If {$\calcrank{ \set{v}_j \cup \{ \vec{q} \}} > \calcrank{ \set{v}_j }$ }	\label{alg1:rank_increase_new}
				\State $\set{Q} := \set{Q} \cup \{ j, \vec{q} \}$	\label{alg1:q_add}
			\EndIf
		\EndFor
	\EndWhile
	\State \Return $\calcrank{ \set{v}_{c} }$	\label{alg1:return}

\EndProcedure
\end{algorithmic}
\caption{Pseudocode for calculating component dimensionality.
\label{alg:graph_search}}
\end{algorithm}
\end{figure}

The input to the algorithm (line \ref{alg1:input}) is a set of graph
edges ($\set{E}$) and a component ($c$) whose dimension we wish to
determine.  We maintain a set of visited or \emph{seen} vertices (line
\ref{alg1:s_init}) and a set of visited vertices for each of the $n$
components in the aperiodic primitive unit cell (line \ref{alg1:v_init}).
A vertex queue is maintained whose elements
consist of a component index and cell coordinates.  The queue is
initialized with the component $c$ in the cell with coordinates $O =
[0, 0, 0]$ (line \ref{alg1:q_init}).
The algorithm runs until the queue is empty (line \ref{alg1:while_q}).
The first element in the queue is extracted and removed (lines
\ref{alg1:q_pop} and \ref{alg1:q_remove}).  If the element has already
been visited it is skipped (line \ref{alg1:seen_check}), otherwise it
is added to the set of visited elements (line \ref{alg1:seen_add}).
If the addition of the vertex serves to increase the rank of the set
of visited vertices (line \ref{alg1:rank_increase}), it is added to
the set (line \ref{alg1:v_add}).

New vertices in adjacent cells are generated from the edge list.  For
a component $i$, the edge list $\set{E}_i$ (line \ref{alg1:edge_loop})
specifies the neighboring components ($j$) and the cell offset
($\vec{\Delta}$), from which the coordinates of the neighboring cell
can be calculated (line \ref{alg1:new_vertex}).
If the neighbor element has already been visited it is either skipped
(line \ref{alg1:seen_check_new}), or added to the queue (line
\ref{alg1:q_add}) if it serves to increase the rank of the set of
visited vertices (line \ref{alg1:rank_increase_new}).
When the queue is empty, the rank of the component is returned (line
\ref{alg1:return}).

\subsection{Interval Identification}
\label{sec:interval_algorithm}

The purpose of the modified method (described in
Algorithm~\ref{alg:incremental}) is to identify intervals in $k$ in
which the dimensionality classification is constant.

\begin{figure}
\begin{algorithm}[H]
\begin{algorithmic}[1]
\Procedure{FindIntervals}{$\set{E}$}	\label{alg2:args}
	\State $k_\text{prev} := 0$	\label{alg2:kprev}
	\State $\set{h}_\text{prev} := \left[ n_\text{atoms}, 0, 0, 0 \right]$.	\label{alg2:hprev}
	\State $\set{R} = \emptyset$		\label{alg2:results}
	\For {$(k, i, j) \in \set{E}$}	\label{alg2:for}
		\State Add edge between vertices $i$ and $j$	\label{alg2:edge}
		\State Identify connected components	\label{alg2:connected}
		\State Update $\set{h}$	\label{alg2:hist} 
		\If{$\set{h} \neq \set{h}_\text{prev}$}	\label{alg2:test}
			\State $\set{R} := \set{R} \cup \left\{ \left(k_{\text{prev}}, k, \set{h}_\text{prev} \right) \right\}$	\label{alg2:addinterval}
		\EndIf
		\If{$\set{h} = \left[0, 0, 0, 1\right]$}	\label{alg2:testfinish}
			\State\Return $\set{R} \cup \left\{ \left( k, \infty, \set{h} \right) \right\}$	\label{alg2:return}
		\EndIf
		\State $k_\text{prev} := k$	\label{alg2:updatekprev}
		\State $\set{h}_\text{prev} := \set{h}$	\label{alg2:updatehprev}
	\EndFor
\EndProcedure
\end{algorithmic}
\caption{Pseudocode for finding all dimensionality intervals.
\label{alg:incremental}}
\end{algorithm}
\end{figure}

The input to the algorithm (line \ref{alg2:args}) is the set of all
possible edges, sorted according to their $k$ values, from lowest to
highest.  Each element in this set, $(k, i, j) \in \set{E}$, contains
the $k$ value of the edge and the indices, $i$ and $j$, of the
vertices it connects.  Periodic boundary conditions must be taken into
account when generating this set.  Due to the periodicity this set is
infinitely large; the relevant (finite) subset, however, can be
generated incrementally.

The algorithm proceeds by inserting edges from $\set{E}$ into the
graph, one by one (line \ref{alg2:for}).  For every edge in the
primitive cell, the corresponding number of edges are inserted into
the supercell.  Connected components in both the primitive cell and
the supercell are identified after each edge insertion (line
\ref{alg2:edge}), from which a \emph{component histogram} is
calculated (line \ref{alg2:hist}).  The histogram, $\set{h}$, contains
the number of 0D, 1D, 2D, and 3D components present.  Prior to any
edge insertion, only 0D components are present, which is reflected in
the initial state of the histogram (line \ref{alg2:hprev}).  If an
edge insertion produces a change in the component histogram (line
\ref{alg2:test}), the $k$ interval is added (line
\ref{alg2:addinterval}) to the set of results (line
\ref{alg2:results}).
The algorithm terminates when the histogram consists only a single 3D
component (line \ref{alg2:testfinish}).  A 3D interval is added to the
results (line \ref{alg2:return}), which implicity contains all
uninserted edges in $\set{E}$: once the dimensionality is fully 3D, no
further edge insertions can change the classification.

The algorithms developed here are included in the ASE~\cite{Larsen:2017hn} library.

\begin{acknowledgments}
  The authors thank Nicolas Mounet and Nicola Marzari for kindly
  providing data for the layered compounds identified
  by~\citet{Mounet:2018ks}, FIZ Karlsruhe -- Leibniz Institute for
  Information Infrastructure for providing CIF files of all entries in
  the ICSD, and anonymous referees for comments which improved the manuscript.
  This work was supported by Grant No. 7026-00126B from the Danish Council for
  Independent Research and by the VILLUM Center for Science of Sustainable Fuels
  and Chemicals which is funded by the VILLUM Fonden research grant (9455).
\end{acknowledgments}

\bibliography{refs}

\end{document}